\documentclass[a4paper,11pt]{article}
\pdfoutput=1 
\usepackage{jheppub} 

\usepackage[T1]{fontenc}  

\usepackage{graphicx}
\graphicspath{{figures/}}

\usepackage{dsfont}
\usepackage{tabularx}
\usepackage{xcolor}
\usepackage{soul}
\usepackage{lipsum}
\usepackage{mathtools}

\usepackage{tikz}

\usepackage{amssymb}
\usepackage{amsfonts}
\usepackage{amsmath}
\usepackage{bm}
\usepackage{cancel}

\usepackage{xspace}
\usepackage{fancyvrb}
\usepackage{longtable}
\usepackage{enumitem}
\usepackage{url}
\usepackage{blindtext}
\usepackage{ulem}

\usepackage[toc,page]{appendix}
\newcommand{\smodels}{\textsc{SModelS}}
\newcommand{\smo}{\textsc{SModelS}\xspace}

\newcommand{\protomodel}{proto-model\xspace}
\newcommand{\Protomodel}{Proto-model\xspace}
\newcommand{\protomodels}{proto-models\xspace}

\newcommand{\M}{\mathbf{M}}

\newcommand{\K}{\ensuremath{K}\xspace}
\newcommand{\SM}{\ensuremath{\mathbf{SM}}\xspace}
\newcommand{\BSM}{\ensuremath{\mathbf{BSM}}\xspace}

\def\met{$E_T^{\rm miss}$}

\title{Artificial Proto-Modelling: Building Precursors of a Next Standard
Model from Simplified Model Results}

\author[a,b,1]{Wolfgang Waltenberger,\note{Corresponding author.}}
\author[c]{Andr\'e Lessa,}
\author[d]{Sabine Kraml}

\affiliation[a]{Institut f\"ur Hochenergiephysik,  \"Osterreichische Akademie
der  Wissenschaften,\\ Nikolsdorfer Gasse 18, A-1050 Wien, Austria}
\affiliation[b]{University of Vienna, Faculty of Physics, Boltzmanngasse 5,
A-1090 Wien, Austria}
\affiliation[c]{Centro de Ci\^{e}ncias Naturais e Humanas, Universidade
Federal do ABC,\\ Santo Andr\'e, SP - Brasil}
\affiliation[d]{Laboratoire de Physique Subatomique et de Cosmologie,
Universit\'e Grenoble-Alpes, CNRS/IN2P3, 53 Avenue des Martyrs, F-38026
Grenoble, France}

\emailAdd{wolfgang.waltenberger@oeaw.ac.at}
\emailAdd{andre.lessa@ufabc.edu.br}
\emailAdd{sabine.kraml@lpsc.in2p3.fr}

\abstract{
We present a novel algorithm to identify potential dispersed signals of new physics in the slew of published LHC results. It employs a random walk algorithm to introduce sets of new particles, dubbed ``\protomodels'', which are tested against simplified-model results from ATLAS and CMS (exploiting the \smo\ software framework). 
A combinatorial algorithm identifies the set of analyses and/or signal regions that maximally violates the SM hypothesis, while remaining compatible with the entirety of LHC constraints in our database. 
Demonstrating our method by running over the experimental results in the \smo\ database, we find as currently best-performing \protomodel a top partner, a light-flavor quark partner, and a lightest neutral new particle with masses of the order of 
1.2~TeV, 700~GeV and 160 GeV, respectively.
The corresponding global $p$-value for the SM hypothesis is $p_\mathrm{global} \approx 0.19$; by construction no look-elsewhere effect applies. 
}

\begin{document}
\maketitle
\flushbottom

\begin{keyword} 
~LHC; supersymmetry; inverse problem; simplified models; 
physics beyond the standard model; reinterpretation
\end{keyword}

\clearpage

\section{Introduction} \label{sec:intro}

Inverse problems~\cite{inverse_problem} are defined to be the process of inferring causal factors and general rules from observational data. They are ubiquitous in all sciences and notoriously hard to tackle. 
In particle physics, we typically refer to the problem of constructing the fundamental Lagrangian from our observations as our field's inverse problem. 
In the context of the Standard Model,
the precise theoretical predictions 
essentially turn this problem into classical hypothesis tests, cf.\ 
for instance the determination of the properties of the Higgs boson. 

Should experimental evidence for physics beyond the Standard Model (BSM) arise at the LHC or elsewhere,  
this mapping from signature space to the parameter space of the underlying theory will, however, be much less clear-cut~\cite{Binetruy:2003cy,ArkaniHamed:2005px}; see also \cite{Knuteson:2006ha,Berger:2007ut,Altunkaynak:2008ry,Bornhauser:2012iy,Gainer:2015zna}. 
The reasons are twofold. First, there is the large, and still growing, number of proposed extensions of the Standard Model (see, e.g., \cite{Lykken:2010mc,ArkaniHamed:2005px} for concise overviews), many of which have a large number of free parameters and can come in multiple, non-minimal variants. Second, very likely not enough information (i.e., too few observables) 
will be available for a direct connection between experiment and theory, making it necessary to ``relate incomplete data to incomplete theory''~\cite{Binetruy:2003cy}. 

At the same time, the LHC has been providing new experimental results at an enormous pace, making it a highly non-trivial task to determine which theories or scenarios 
survive experimental scrutiny and which do not.
Testing a BSM model or inferring its preferred region of parameter space typically requires the construction of a likelihood, which  encapsulates the information from the relevant LHC results and/or from other experiments.
Building such a likelihood can be a daunting task, due to the limited amount of information available outside the experimental collaborations\footnote{The situation has improved drastically in the last couple of years, cf.\ the discussion in  \cite{Abdallah:2020pec,Kraml:2012sg}.}
and the computational resources required for exploring large parameter spaces. 
Thus, even when possible, the statistical inference obtained
is naturally limited to the concrete BSM model under investigation.
Owing to the vast number of proposed models (and new ones still to come) it is not a given that this top-down approach will direct us towards the hypothetical Next Standard Model (NSM)~\cite{ArkaniHamed:2007fw}.

In this work, we therefore aim to tackle the inference of the NSM through a bottom-up approach, which strongly relies on LHC data and contains only minimal theoretical bias. 
The idea itself is not new: systematic, bottom-up approaches to the LHC inverse problem were envisaged previously 
with the \textsc{Bard} algorithm in~\cite{Knuteson:2006ha},  
the \textsc{MARMOSET} framework in~\cite{ArkaniHamed:2007fw}, 
and the characterisation of new physics through simplified models in~\cite{Alwall:2008ag}. 
Our approach is different in the sense that, given the absence of any clear signal of new physics so far, we focus on potential dispersed signals that may have been missed in the common analysis-by-analysis interpretations of the data. 

\begin{figure}[t!]
\begin{center}
\includegraphics[width=0.9\textwidth]{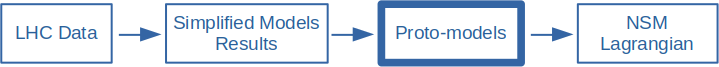}
\vspace*{4mm}
\caption{Overall strategy of how we may envisage to construct an NSM from LHC data:
the outcomes of specific searches, which are communicated via simplified model results, are used to construct \protomodels. These may be scrutinized in dedicated analyses and eventually help to infer the NSM; see also \cite{Knuteson:2006ha,ArkaniHamed:2007fw,Alwall:2008ag}.}
\label{fig_workflow}
\end{center}
\end{figure}

Concretely, we introduce so-called ``\protomodels'', which can be viewed as sets
or ``stacks'' of simplified models, where the number of BSM particles as well as their masses, production cross sections and decay branching ratios are taken as free parameters.
Through the use of simplified-model constraints via the 
\smodels~\cite{Kraml:2013mwa} package, we construct approximate
likelihoods for the \protomodels, allowing us to perform Markov Chain Monte Carlo (MCMC)-type walks in their parameter space.  
These MCMC walks consist of random changes in the BSM number of
particles, their masses, production cross sections and decay branching ratios, with the goal of 
finding \protomodels which evade all available constraints and at the
same time explain potential dispersed signals in the data. 
We stress that \protomodels are not intended to be UV complete nor
to be a consistent effective field theory. Nonetheless we hope they 
can be useful to guide future experimental and theoretical efforts, 
and perhaps even serve 
as a first step toward a construction of the NSM, as illustrated in
Fig.~\ref{fig_workflow}.

The rest of this paper is organized as follows.
Section~\ref{sec:protomodels} discusses in detail our definition of \protomodels, their parameters and the assumptions they are based on. 
Section~\ref{sec:predictions} describes the method used for computing the LHC constraints for a given \protomodel as well as how we use this information to construct an approximate likelihood.
The MCMC walk and the algorithm used for building new \protomodels is detailed in Section~\ref{sec:walker}. 
Finally, in Section~\ref{sec:results}, we apply the MCMC walk to the full \smo database and discuss the results obtained. 
We conclude in Section~\ref{sec:conclusions} with a short summary and charting out future developments and extensions of this work. 
A set of appendices provides complementary information on 
the MCMC-type walk  over  the \protomodels   
parameter  space (Appendix~\ref{sec:builder}),  
posterior densities of \protomodel parameters and 
distributions of the test statistic (Appendix~\ref{sec:distributions}) and the simplified-model 
results included in the \smo v1.2.4 database (Appendix~\ref{sec:anatables}).

\section{Proto-Models} \label{sec:protomodels}

As mentioned in the Introduction, \protomodels are defined by their BSM particle content, the particle masses, production cross sections and decay branching ratios. In this work, we assume that all particles either decay promptly or are fully stable at detector scales. 
Since \protomodels are not intended to be fully consistent theoretical models, their properties are   
not bound by higher-level theoretical assumptions, such as representations of the SM gauge groups or higher symmetries. 
Nonetheless, since we will make extensive use of simplified-model results from searches for supersymmetry (SUSY) 
we impose the following constraints:
\begin{enumerate}
    \item all BSM particles are odd under a $\mathcal{Z}_2$-type symmetry, so they are always pair produced and always cascade decay to the lightest state;
    \item the Lightest BSM Particle (LBP) is stable and electrically and color neutral, and hence is a dark matter candidate;
    \item except for the LBP, all particles are assumed to decay promptly; 
    \item only particles with masses within LHC reach are considered part of a specific \protomodel.
\end{enumerate}

In the current version of the algorithm, we allow \protomodels to consist of up to 20 BSM particles as described in the next subsection. 

\subsection{Particle Spectrum}
\label{ssec:particlecontent}

Unlike ``full'' models, the BSM particle content of a \protomodel is not fixed.
Since the number of degrees of freedom (including spin and multiplicity of states) for each particle is mostly relevant for the production cross section, which is treated as a free parameter, we do not specify the spin or multiplicity of the BSM particles. 
In the current version, we consider the following pool of 20 (SUSY inspired) particles:

\begin{itemize}

\item Light quark partners $X_{q}$ ($q=u,d,c,s$): we allow for a single partner for each light-flavor quark. Unlike SUSY models, we do not consider two independent particles (left and right-handed squarks) for each flavor. As mentioned above, a possible multiplicity of (degenerate) states is accounted for by rescaling the production cross section.

\item Heavy quark partners $X_{b}^{i}$,~$X_{t}^{i}$ ($i=1,2$): unlike the light quark partners, we consider two independent particles for each flavor. Since these particles have so extensively been searched for at the LHC, we include two states in order for the \protomodels to have enough degrees of freedom to accommodate the data.

\item Gluon partner $X_{g}$: we introduce one new color-octet particle, 
analogous to a gluino in SUSY. 

\item Electroweak partners $X_{W}^{i}$,~$X_{Z}^j$ ($i=1,2$; $j=1,2,3$): we allow for two electrically charged and three neutral states. These might correspond to charginos and neutralinos in the MSSM (with the neutral higgsinos being exactly mass-degenerate), or to the scalars of an extended Higgs sector with a new conserved parity. The lightest neutral state ($X_Z^1$) is assumed to be the LBP. 

\item Charged lepton partners $X_\ell$ ($\ell=e,\mu,\tau$): like for light-flavor quarks, we consider a single partner for each lepton flavor.

\item Neutrino partners $X_{\nu_\ell}$: again we consider one partner for each neutrino flavor ($\nu_e,\nu_\mu,\nu_\tau$).

\end{itemize}

As mentioned above, the lightest BSM particle is required to be the  $X_{Z}^{1}$, hence all masses must satisfy $m(X) \ge m(X_{Z}^{1})$. In a given \protomodel only particle masses below 2.4~TeV are considered, so they are within the LHC reach. States which appear in two- or three-fold multiplicities are mass ordered, e.g., $m({X_t^2})>m({X_t^1})$. 

Some additional requirements are necessary for the masses of the colored new particles, to avoid that our machinery ``discovers'' light states in regions which are poorly covered in the database. Concretely, for the $X_{g}$ and $X_{q}$ we disallow masses below 310~GeV, since most of the simplified-model results we employ do not consider masses below this value.
Also, for $X_{t}$ masses below 280 GeV, we disallow the region $150\;\mathrm{GeV} < m(X_{t})-m(X_{Z}^{1}) < 200\;\mathrm{GeV}$. The reason is that many CMS analyses do not make any statements for this ``corridor'' region, as events in there become too similar to $t \bar{t}$ events, see e.g.\ Ref.~\cite{Sirunyan:2019ctn}.

\subsection{Decay Channels and Branching Ratios}

The decay modes of each new particle must of course be consistent with its quantum numbers. At present, we restrict the BSM particle decays to the channels shown in Table~\ref{tab:decays}. Note that not all possibilities are considered in this work. In particular 
$X_W$ and $X_Z$ decays into $X_\ell+\ell/\nu$, $X_\nu+\nu/\ell$, and $X_{W,Z}+\gamma$ are not yet taken into account.

Specific decay modes are turned off if one of the daughter particles is not present in the \protomodel,  or if the decay is kinematically forbidden. The branching ratios (BRs) of the allowed channels are taken as free parameters, which however must add up to unity.

\begin{table}
	\setlength\extrarowheight{3pt}
	\centering
	\begin{tabular}{c|c||c|c}
		particle & decay channels & particle & decay channels\\
		\hline
		$X_{q}$ & $q X_{Z}^{j},\; q'  X_{W}^{i},\; q X_{g}$ 
		& $X_{W}^{1}$ & $W X_{Z}^{j}$\\ 
		\hline 
		$X_{t}^{1}$ & $t X_{Z}^{j},\; b X_{W}^{i},\; W X_{b}^{1},\;tX_{g}$
		& $X_{W}^{2}$ & $W  X_{Z}^{j},\; Z X_{W}^{1}, \; h X_{W}^{1}$\\ 
		\hline 
		$X_{b}^{1}$ & $b X_{Z}^{j},\; t X_{W}^{i},\; W X_{t}^{1},\; b X_{g}$
		& $X_{Z}^{j\not=1}$ & $W  X_{W}^{i},\; Z X_{Z}^{k}, \; h X_{Z}^{k}$\quad \\ 
		\hline
		$X_t^2$ & $ tX_Z^j,\; bX_W^i,\; ZX_t^1,\; WX_b^1,\; tX_g $ 
		& $X_\ell$ & $\ell X_{Z}^{j},\; \nu_\ell  X_{W}^{i}$ \\ 
		\hline
		$X_b^2$ & $ bX_Z^j,\; tX_W^i,\; ZX_b^1,\; WX_t^1,\; bX_g$   
		& $X_{\nu_\ell}$ & $\nu_\ell  X_{Z}^{j},\; \ell X_{W}^{i}$ \\ 
		\hline
		$X_{g}$ & \multicolumn{3}{l}{ $q\bar{q} X_Z^i,\; q\bar{q}' X_W^i,\; b\bar{b} X_Z^i,\; t\bar{t} X_Z^j,\; bt X_W^i,\; qX_q,\; bX_b,\; tX_t $} \\ \hline
	\end{tabular}
	\caption{The 20 BSM states and their decay channels considered for constructing \protomodels in this work. 
}
	\label{tab:decays}
\end{table}

\subsection{Production Cross Sections}
\label{sec:xsecs}

In addition to the masses and BRs of the BSM particles, their production cross sections are also allowed to vary freely. 
However, in order to have sensible starting values, the cross sections are first computed assuming the BSM particles to be MSSM-like, and then allowed to be rescaled freely by {\it signal strength multipliers} $\kappa$. 
For instance, the pair production of $X_{g}$ is given by:
\begin{equation}
 \sigma\left(p p \to X_{g} X_{g}\right) = \kappa_{X_g X_g} \times \sigma\left( p p \to \tilde{g} \tilde{g}\right)
\end{equation}
with the mass of the gluino $\tilde{g}$ set to the $X_g$ mass. The rescaling factors $\kappa_{X_iX_j}$ 
are taken as free parameters of the \protomodel. 

In practice, we use a template SLHA input file for the MSSM to define masses and BRs of the \protomodel (with $\tilde q_R$, $\tilde\chi^0_4$ and heavy Higgses decoupled); the reference SUSY cross sections are then computed with Pythia 8.2~\cite{Sjostrand:2014zea} and NLLFast 3.1~\cite{Beenakker:1996ch,Kulesza:2008jb,Kulesza:2009kq,Beenakker:2009ha,Beenakker:2011fu,Beenakker:1997ut,Beenakker:2010nq}. 
We note that this way cross sections which are naturally suppressed or theoretically not allowed (such as $p p \to X_e^- X_e^-$) are automatically neglected. While this adds a certain bias, as processes which do not occur in the SUSY case (like production of two different top partners, $pp\to X_t^1\bar X_t^2$) will also be absent in the \protomodel, this is not a problem, because there are currently no simplified-model results available for such cases.

\section{LHC Results}
\label{sec:predictions}

In order to quickly confront individual \protomodels with a large number of LHC results, we make use of the  \smo~\cite{Kraml:2013mwa,Ambrogi:2017neo,Dutta:2018ioj,Heisig:2018kfq,Ambrogi:2018ujg,Khosa:2020zar,Alguero:2020grj} software package. 
This allows us to evaluate LHC constraints for a given input model without the need of Monte Carlo simulation.
The experimental results stored in the \smo database fall into two main categories:
\begin{itemize}
	\item {\bf Upper Limit (UL)} results: these correspond to the observed limits on the production cross section (times BR) for simplified model topologies as a function of the BSM masses obtained by the experimental collaborations, $\sigma^\mathrm{UL}_\mathrm{obs}$. Some results also include the expected upper limits, denoted as $\sigma^\mathrm{UL}_\mathrm{exp}$. All limits are given at 95\% confidence level (CL).
	\item {\bf Efficiency Map (EM)} results: these correspond to signal efficiencies (more precisely, acceptance $\times$ efficiency, ${\cal{A}}\times \epsilon$, values) for simplified topologies as a function of the BSM masses for the  signal regions considered by the corresponding experimental analysis. These results also include information about the number of observed and expected events (incl.\ uncertainties thereon) for each signal region. 
\end{itemize}

While providing conservative results, the \smo tool is computationally cheap, making it feasible to test hundreds of thousands of \protomodels within reasonable time. The current \smo database v1.2.4~\cite{Dutta:2018ioj,Khosa:2020zar,Alguero:2020grj} includes results from 40 ATLAS and 46 CMS experimental searches for a variety of final states with missing transverse energy (\met) 
at $\sqrt{s}=8$ and 13~TeV, corresponding to about 250 upper limit maps and  1,700 individual efficiency maps. 
The included analyses and types of results are detailed in Appendix~\ref{sec:anatables}. 
This large database renders a global approach to scrutinizing the LHC results for possible dispersed signals feasible. 

Key to this is the construction of an approximate likelihood for the signal, 
\begin{equation}
    \mathrm{L}_{\BSM}(\mu|D) =  P\left(D|\mu + b + \theta \right) p(\theta) \,,
\end{equation}
which describes the plausibility of the signal strength $\mu$, given the data
$D$.
Here, $\theta$ denotes the nuisance parameters describing systematic
uncertainties in the signal ($\mu$) and background ($b$), while $p(\theta)$
corresponds to their probability distribution function.\footnote{Note that $\mu$ is a {\it global} signal strength.
Therefore the signal cross sections are given by $\sigma = \mu \times \kappa \times \sigma_{\mathrm{SUSY}}$, where $\kappa$ are the signal strength multipliers defined in Section~\ref{sec:xsecs}.  Technically, once the value of $\mu$ which maximizes the likelihood ($\hat{\mu}$) is determined, all signal strength multipliers are rescaled by $\kappa \to \hat{\mu} \times \kappa$ and $\mu$ is taken as 1.}
In the following two subsections, we explain the two main steps to arrive at $\mathrm{L}_{\BSM}$: 
computing the individual likelihoods for each applicable  analysis, and  combining the individual likelihoods into a global one.

\subsection{Likelihoods and Constraints from Individual Analyses}\label{sec:likelihoods}

The extent to which likelihoods can be computed crucially depends on the information available from the experimental collaboration.

\begin{enumerate}

    \item If only observed ULs are available, the likelihood becomes a constraint in the form of a step function at the observed 95\% CL exclusion limit. This is in fact not useful for constructing $\mathrm{L}_{\BSM}$ per se, but will be used to determine the maximal allowed signal strength $\mu_{max}$ (see Section~\ref{sec:combining} below).
    
    \item  If the expected ULs are available in addition to the observed ones, following~\cite{Azatov:2012bz} we approximate the likelihood as a truncated Gaussian:
    \begin{equation}
       \mathrm{L}(\mu|D) =\frac{c}{\sqrt{2\pi}} \frac{\sigma_\mathrm{ref}}{{\bm \sigma}_\mathrm{obs}}\,e^{-(\mu \sigma_\mathrm{ref}-\sigma_\mathrm{max})^2/2{\bm \sigma}^2_\mathrm{obs}},~\mathrm{for}~ \mu \ge 0\,. \label{eq:llhfromlimit}
    \end{equation} 
    Here,  the likelihood is a function of the signal strength multiplier $\mu$ for any reference cross section $\sigma_\mathrm{ref}$ 
    (in our case the cross section predicted for the given \protomodel), while 
    ${\bm\sigma}_\mathrm{obs}$ is approximated
    with the standard deviation of the {\it expected} Gaussian likelihood, 
    ${\bm\sigma}_\mathrm{obs} \approx {\bm\sigma}_\mathrm{exp} = \frac{\sigma_\mathrm{exp}^{\mathrm{UL}}}{1.96}$, with  $\sigma_\mathrm{exp}^{\mathrm{UL}}$ the expected 95\% CL upper limit on the signal cross section. 
    In addition, $\sigma_\mathrm{max}$ is chosen such that the approximate truncated Gaussian likelihood correctly reproduces the 95\% CL observed limit on the production cross section, $\sigma_\mathrm{obs}^{\mathrm{UL}}$\,. Finally, $c$ is a normalization constant, $c \ge 1$, which ensures that the integral over the likelihood from $\mu=0$ to infinity equals unity. 
    While Eq.~\eqref{eq:llhfromlimit} is certainly a crude   approximation in some cases, it does contain valuable information on possible excesses in the data.
    
    However, we cannot expect Eq.~\eqref{eq:llhfromlimit} to hold for large excesses. Therefore, if the observed UL differs from the expected one by more than two standard deviations, we ``cap'' the likelihood by artificially setting the observed UL to 
    $\sigma_\mathrm{obs}^{\mathrm{UL}} = \sigma_\mathrm{exp}^{\mathrm{UL}} + 2 \bm\sigma_\mathrm{exp}$.
    With this procedure we avoid overly optimistic interpretations of excesses.
    While this affects only a small fraction of results, 
    this is a crude approximation, and it could be avoided by EM results, see below.  

    \item EM results contain more information to construct a proper likelihood. 
In the absence of a full statistical model (see point 4 below), we assume $p(\theta)$ to follow a Gaussian distribution centered
around zero and with a variance of $\delta^2$,
whereas $P(D)$ 
corresponds to a counting variable and is thus 
properly described by a Poissonian. The likelihood is then given by 
\begin{equation}
	\mathrm{L}(\mu|D) = \frac{(\mu s + b + \theta)^{n_{obs}} e^{-(\mu s + b +
\theta)}}{n_{obs}!} {\rm exp} \left( -\frac{\theta^2}{2\delta^2} \right) \,,
\end{equation}
where $n_{obs}$ is the number of observed events in the signal region under consideration, $b$ is the number of expected background events, and 
$\delta^2=\delta_b^2+\delta_s^2$ the signal+background uncertainty. 
The nuisances $\theta$ can be profiled or marginalized over; our default is profiling.  
This is often referred to as a {\it simplified likelihood}~\cite{Fichet:2016gvx,simplifiedlikelihoods,Buckley:2018vdr}.
While $n_{obs}$, $b$ and $\delta_b$ are directly taken from the experimental publications, we assume a default 20\% systematical uncertainty on the signal. See \cite{Ambrogi:2017neo} for more detail. 

Generally, in each analysis we use only the signal region with the highest $\sigma_\mathrm{obs}^\mathrm{UL} / \sigma_\mathrm{exp}^\mathrm{UL}$ ratio. A large ratio indicates an excess has been observed and allows us to identify potential dispersed signals. Although 
it is not a given that this is the best criterion when combining results from distinct analyses, it allows us to drastically reduce the number of possible signal region combinations across analyses.

\item The best case is to have EM results together with a statistical model, which describes the correlations of uncertainties across signal regions~\cite{Kraml:2012sg,Abdallah:2020pec}. 
In this context, CMS sometimes provides covariance matrices, which  allow for the combination of signal regions as discussed in \cite{simplifiedlikelihoods}. (N.b., this is still a simplified likelihood assuming Gaussian uncertainties.) 
So far, however, 
a covariance matrix together with simplified model EMs is available for only one analysis~\cite{CMS-PAS-SUS-16-052}.

ATLAS, on the other hand, has recently started to publish  
{\it full likelihoods} using a plain-text (JSON) serialization~\cite{ATL-PHYS-PUB-2019-029} of the likelihood, which describes statistical and systematic uncertainties, and their correlations across signal regions, at the same fidelity as used in the experiment.\footnote{This JSON format describes the HistFactory family of statistical models~\cite{Cranmer:1456844}, used by the majority of ATLAS searches; 
it can conveniently be used via the {\tt pyhf} package~\cite{pyhfsw} in a scientific Python environment.}
At present, EM results 
with JSON likelihoods are available in the \smo database for three ATLAS SUSY analyses \cite{Aad:2019byo,Aad:2019pfy,Aad:2019vvf} 
for full Run~2 luminosity (139~fb$^{-1}$), leading to clear improvements in the statistical evaluation for these analyses~\cite{Alguero:2020grj}.%
\footnote{While, as we will see, \cite{Aad:2019byo,Aad:2019pfy,Aad:2019vvf} do not impact our results, we stress that full likelihoods are a boon for reinterpretation studies. Once available on a broader basis, they will be extremely beneficial in particular for global analyses, including the one presented in this paper.}

\end{enumerate}

\subsection{Building a Global Likelihood}
\label{sec:combining}

If we had perfect knowledge of the LHC results and their mutual correlations, we would simply compute a single joint likelihood for all results. 
However, cross-analysis correlations are very difficult to quantify and require a dedicated effort by the experimental collaborations. Examples are the combinations of Higgs measurements~\cite{Sirunyan:2018koj,Aad:2019mbh}, which do provide channel-by-channel correlations as auxiliary public material.  
For searches, statistical combinations of several analyses (considering different final states) are rare\footnote{One example is the CMS combination of electroweakino searches \cite{Sirunyan:2018ubx}.} and if done, the cross-analysis correlations have not been
made public so far.

In this work, we therefore treat such correlations in a binary way---a given pair of analyses is either considered to be approximately uncorrelated, in which case it may be combined (by multiplying the respective likelihoods), or it is not at all considered for combination.
We always assume results from different LHC runs, and from distinct experiments (ATLAS or CMS) to be approximately uncorrelated. Furthermore, we also treat results with clearly different final states in their signal regions (e.g., fully hadronic final states vs.\ final states with leptons) to be uncorrelated. 

\begin{figure}[t!]
	\begin{center}
		\includegraphics[width=.49\textwidth]{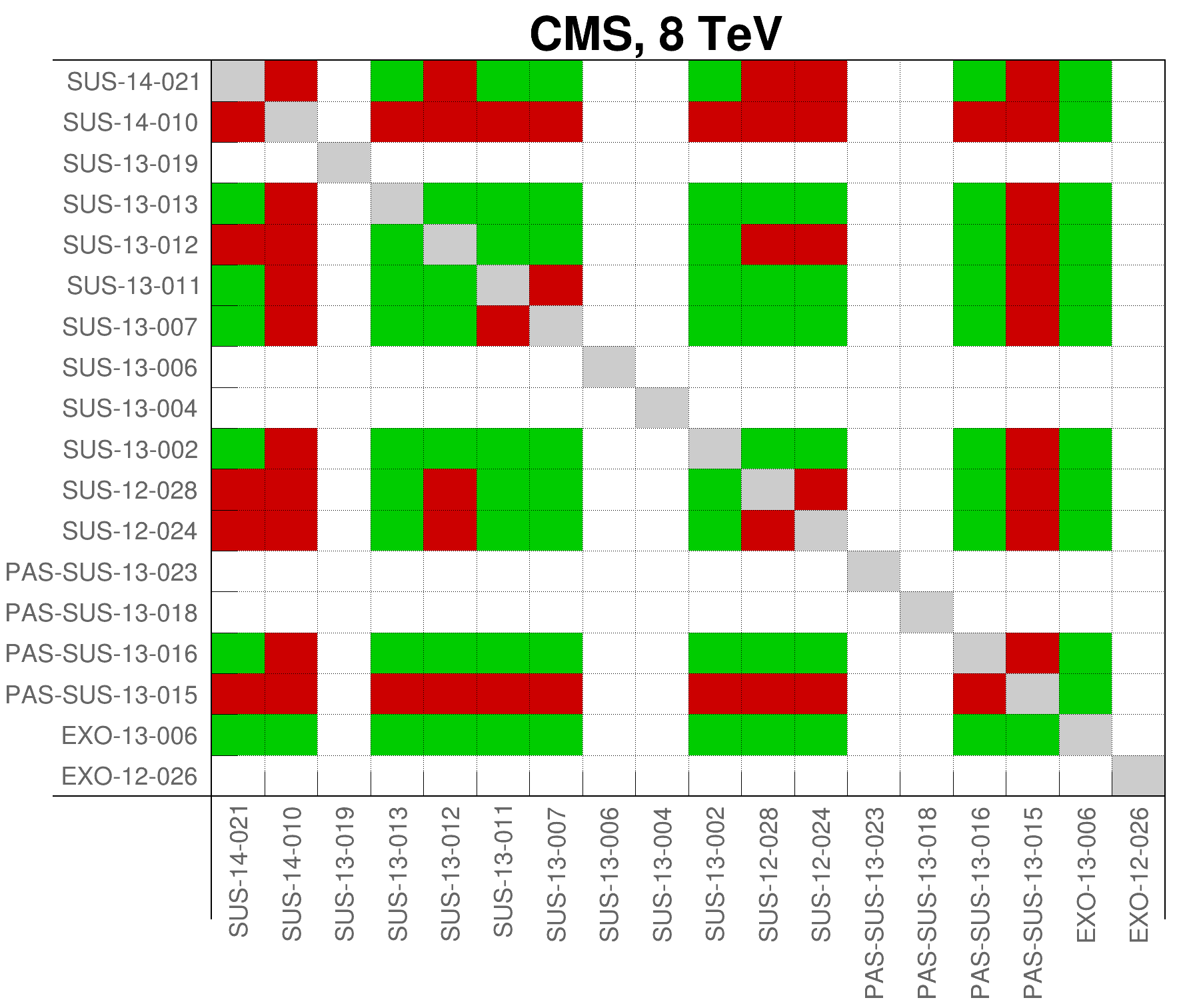}%
		\includegraphics[width=.49\textwidth]{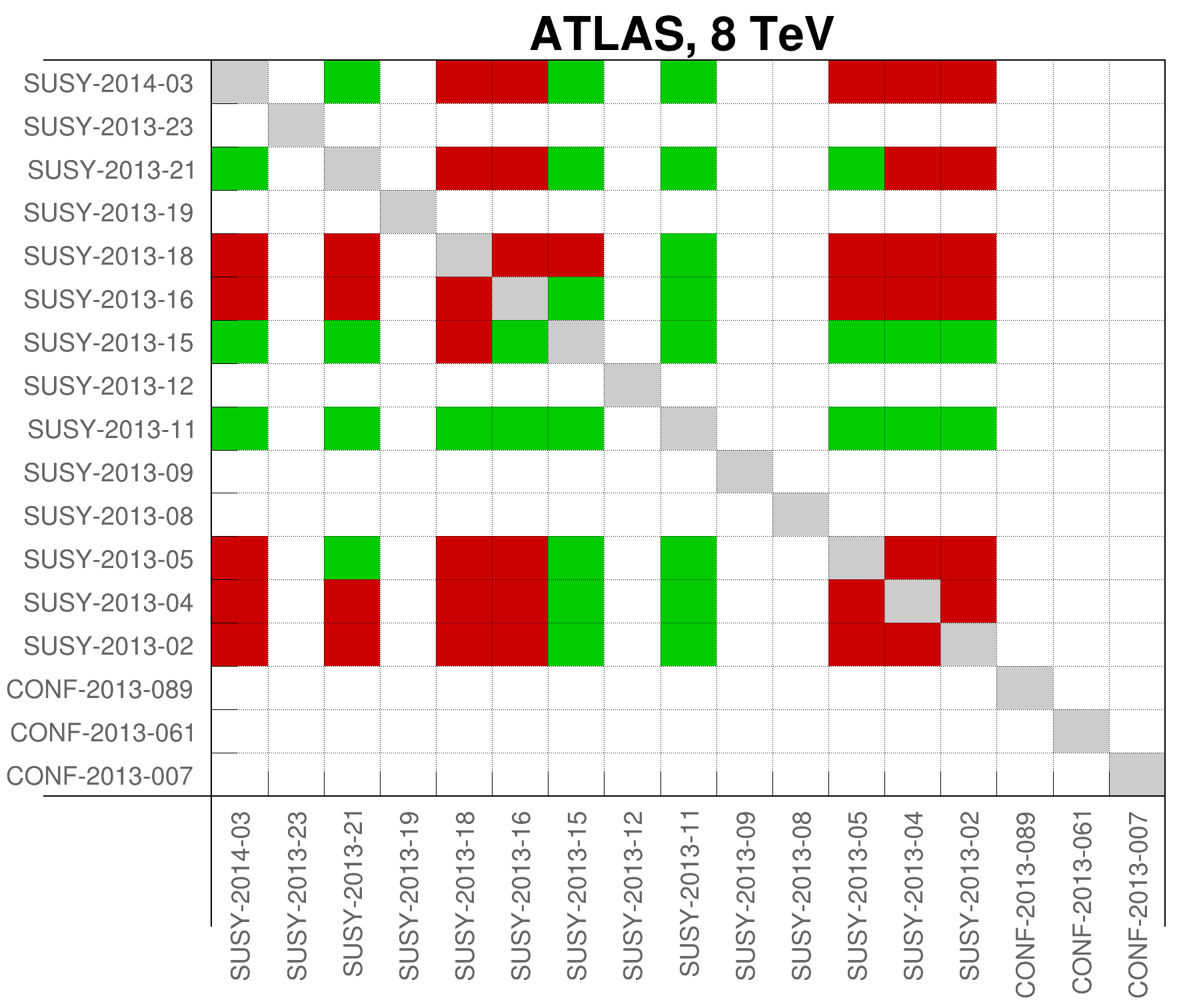}\\[3mm]
		\includegraphics[width=.49\textwidth]{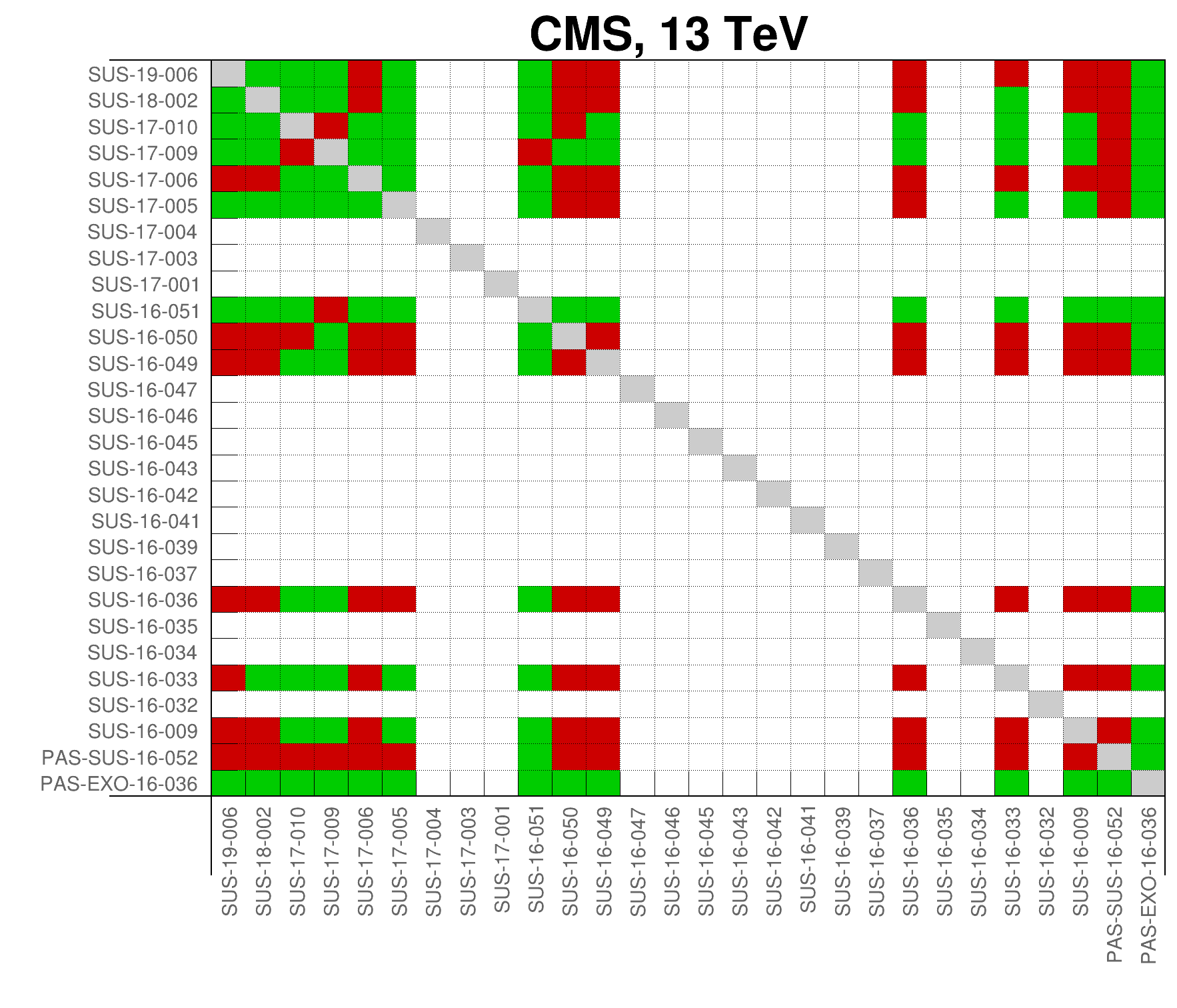}%
		\includegraphics[width=.49\textwidth]{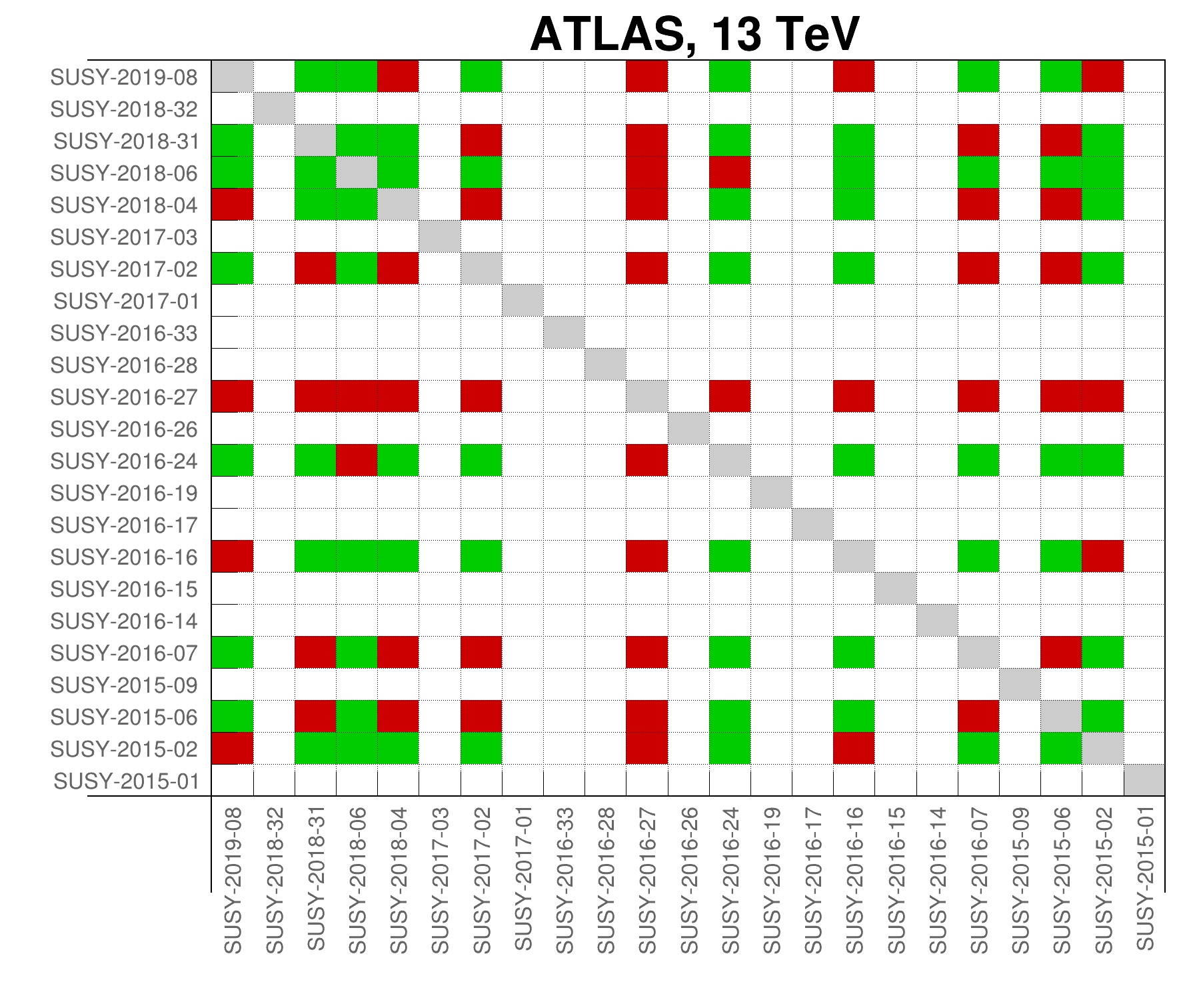}
		\caption{Binary correlation matrices used in this work, describing which analyses are considered
			to be approximately uncorrelated (green bins) and which are not (red bins). Analyses from different LHC runs or different experiments are always taken as uncorrelated. White bins denote analyses, for which no likelihood can be computed (observed UL only).}
		\label{fig_matrix}
	\end{center}
\end{figure}

Figure~\ref{fig_matrix} summarizes our treatment of correlations in the form of four  {\it binary analysis correlation matrices}. We understand that, although well motivated, this is but an educated guess from our side. 
Future runs of our algorithm will be based on more reliable estimates of the results' mutual statistical independence. 
Indeed work on systematically constructing such analyses correlation matrices is underway, see~\cite{leshouches} and contribution~16 in~\cite{Brooijmans:2020yij}. 
However, while a good number of analyses can be combined in our approach, we also see from the white bins in Fig.~\ref{fig_matrix}, that about half (one third) of the 13~(8)~TeV results are observed ULs only, and thus no proper likelihood can be computed for them. 
Nonetheless, under the above assumptions it is possible to construct an approximate combined likelihood for subsets of LHC results. We will refer to each of these subsets as a {\it combination} of results. Each combination must satisfy:
\begin{itemize}
	\item any pair of results in the subset must be considered as uncorrelated and
	\item any result which is allowed to be added to the combination, is in fact added.
\end{itemize}

Likelihoods from uncorrelated analyses can simply be multiplied: 
$\mathrm{L}_{\BSM}(\mu) = \prod_{i=1}^{n} \mathrm{L}_i(\mu)$, where the product is over all $n$ uncorrelated analyses and $\mu$ is the global signal strength. 
Information from all the other analyses, which are not included in the combination, is
accounted for as a constraint on the global signal strength $\mu$.
Given the upper limit on the signal cross section obtained from the most sensitive analysis/signal region, we compute an upper limit on $\mu$: 
\begin{equation}
\mu \times \sigma < 1.3\, \sigma_\mathrm{obs}^\mathrm{UL} \;\Rightarrow\; \mu < \mu_{max} = 1.3 \,\frac{\sigma_\mathrm{obs}^\mathrm{UL}}{\sigma} \;,
\end{equation}
where $\sigma$ is the relevant signal cross section and the 1.3 factor allows for a 30\% violation of the 95\% CL observed limit.
We accept such mild transgressions in order to account for the fact that when simultaneously checking limits from a large number of analyses, a few are statistically allowed to be violated. 
Imposing the above limit on the global signal strength $\mu$ corresponds to truncating the likelihood at $\mu_{max}$ and ensures constraints from correlated analyses, as well as analyses for which no proper likelihood can be computed, are approximately taken into account.

\begin{figure}[t!]
\hspace*{-7mm}\includegraphics[width=0.54\textwidth]{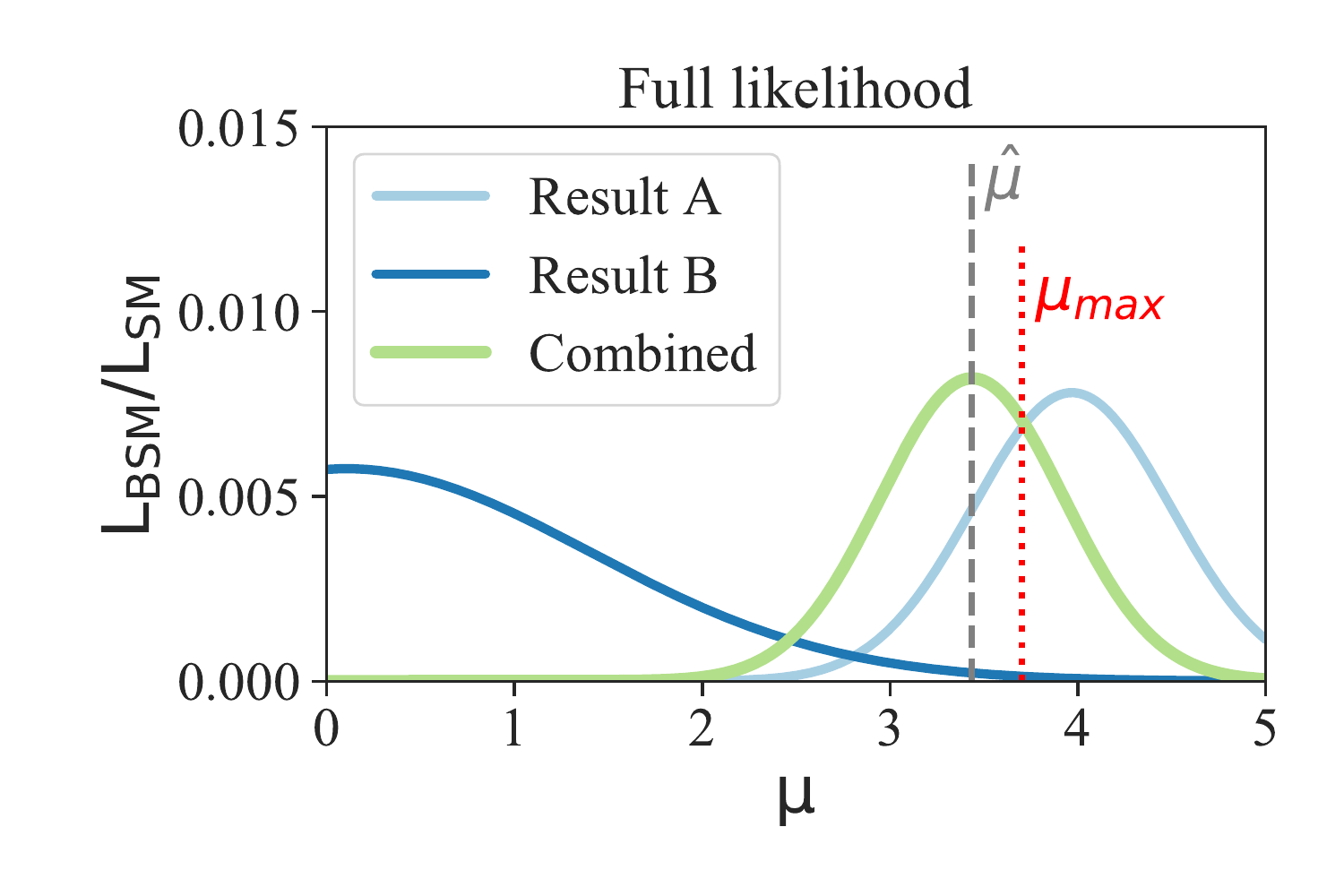}\hspace*{-3mm}\includegraphics[width=0.54\textwidth]{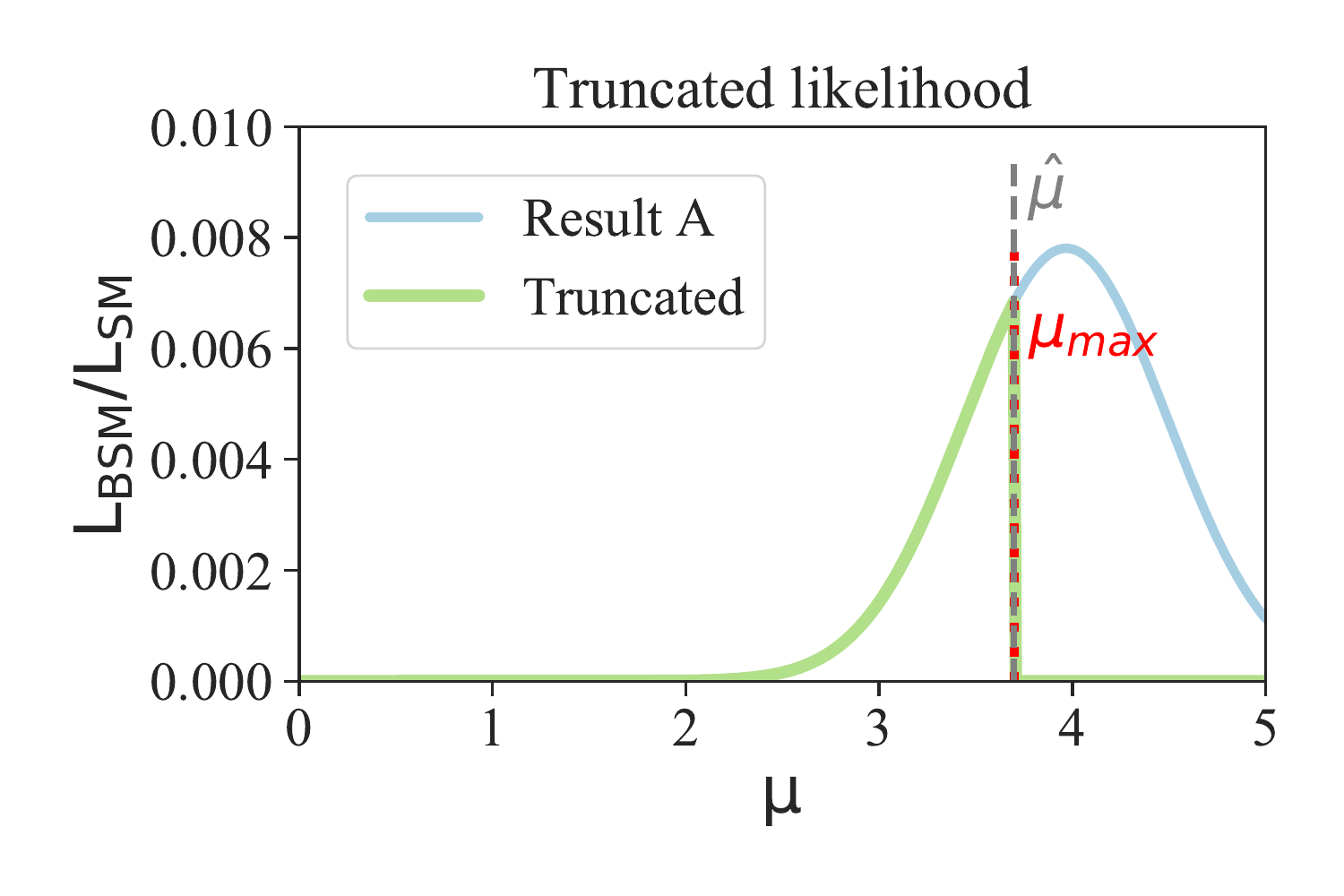}
\vspace*{-10mm}
   \caption{{\it Left}: an illustration of the signal likelihood as a function of the signal strength ($\mu$) for two results (A and B) and the corresponding combined likelihood. The value of $\hat{\mu}$ given by the maximum of the combined likelihood (green curve) is also shown as well as the upper limit on $\mu$ ($\mu_{max}$) given by Result B alone. {\it Right}: a similar example, but where it is not possible to combine both likelihoods, or the likelihood for Result B is not available. In this case the likelihood for Result A is truncated at $\mu_{max}$ (obtained from Result B). All the curves are normalized to the corresponding SM likelihood value ($\mu = 0$).}
   \label{fig_llhd}
\end{figure}

The above procedure is illustrated in Fig.~\ref{fig_llhd}.
The left panel in this figure shows how likelihoods are combined for a \protomodel where it is possible to construct a likelihood for all the relevant results,   
which can all be considered as uncorrelated. In the example shown, ``Result A'' represents an analysis which favors $\mu >0$, while ``Result B'' corresponds to a ``constraining analysis''. The latter observed an under fluctuation in the data with respect to the expected SM background, thus favoring $\mu \leq 0$. The grey line shows the maximum of the combined likelihood ($\hat{\mu}$), while the red line corresponds to the limit on $\mu$ ($\mu_{max}$) obtained from ``Result B'' alone.

The right panel in Fig.~\ref{fig_llhd} displays our approximation for the case where it is not possible to construct   
a proper likelihood for the most constraining analysis (``Result B'' in this example) or it cannot be considered as uncorrelated with the other results.
In this case we truncate the likelihood for ``Result A'' at $\mu_{max}$ obtained from ``Result B''. 
Hence, the value of $\hat{\mu}$ for the truncated likelihood  shifts slightly with respect to the full result. In this work we apply this approximation whenever it is not possible to construct a full likelihood.

\section{The Walker}
\label{sec:walker}

Let us now turn to the algorithm for performing an MCMC-type walk in the \protomodel's parameter space in order to identify
the models that best fit the data. 
This algorithm, which we dub the {\bf walker}, 
is composed of several
building blocks, or ``machines'' that interact with each other in a
well-defined fashion:

\begin{enumerate}
 \item Starting with the Standard Model, a {\bf builder} creates \protomodels,
randomly adding or removing particles and changing any of the \protomodel parameters (see Appendix~\ref{sec:builder} for details).

 \item The \protomodel is then passed on to the {\bf critic} which checks the
model against the database of simplified-model results to determine an upper bound on an overall signal strength ($\mu_{max}$). 
 \item The {\bf combiner} identifies all possible combinations of results and constructs a combined likelihood for each subset (see Section~\ref{sec:combining} for details).
 \item Using the combinations provided by the combiner, the walker computes a test statistic \K for the \protomodel (see Section~\ref{sec:teststatistic} below). If the \K value is higher than the one obtained in the last step, the new \protomodel is kept. If it is lower than the previous \K, the step is reverted with a probability of 
 $\exp\left[\frac{1}{2}(\K_i - \K_{i-1})\right]$,
 where $i$ is the index of the current step.
\end{enumerate}

By many iterations of the above procedure, the walker is able to identify \protomodels which evade all simplified-model limits and at the same time can explain dispersed signals in the data. 
This will be illustrated by means of a toy walk in Section~\ref{sec:toywalk} and by walking over the full \smo database in Section~\ref{sec:results}.

\subsection{Computing a \Protomodel's Test Statistic}
\label{sec:teststatistic}

As mentioned above, during the MCMC-type walk, the algorithm aims to maximize the \protomodel test statistic \K. Naturally, \K must be designed such that it increases for models 
which better satisfy all the constraints (which includes better fitting potential dispersed signals). Furthermore, it is desirable to reduce the test statistic of models with too many degrees of freedom in order to enforce the {\it law of parsimony}, %
that is, Occam's razor~\cite{occam}. Given a \protomodel, we define for each 
combination of results $c \in C$ the auxiliary quantity
\begin{equation}
\K^{c} := 2 \ln \frac{
	\mathrm{L}^{c}_{\BSM}(\hat{\mu}) \cdot \pi(\BSM)}{\mathrm{L}^{c}_{\SM} \cdot \pi(\SM)} \,.
\label{eq:Kc}
\end{equation}

\noindent 
Here $\mathrm{L}^{c}_\BSM$ is the likelihood
for a combination $c$  of experimental results {\it given} the \protomodel, evaluated at the signal strength value $\hat{\mu}$, which maximizes the likelihood and satisfies $0 \leq \hat{\mu} < \mu_{max}$.  $\mathrm{L}^{c}_{\SM}$ is the corresponding SM likelihood, given by  $\mathrm{L}^{c}_\BSM (\mu=0)$. 
Finally, $\pi(\SM)$ and $\pi(\BSM)$ denote respectively the priors for the SM and the \protomodel. The total set of 
combinations of results, $C$, is determined as explained in Section~\ref{sec:predictions}. 
We shall use this auxiliary quantity to define the test statistic \K for a \protomodel simply as
\begin{equation}
 \K := \max\limits_{\forall c \in C} \K^c \,.
 \label{eq:bayesiantest}
\end{equation}

Some explanations are in order regarding the choice of the prior. Since $\pi(\SM)$ is a common factor for all combinations and does not affect the comparison between distinct \protomodels, we wish to define the prior such that 
\begin{equation}
\label{eq:priorsm}
\pi(\SM) := 1 \,.
\end{equation}
Moreover, as mentioned above, the \protomodel prior $\pi(\BSM)$ should penalize the test statistic for newly introduced particles, branching ratios, or signal strength multipliers. We therefore heuristically choose the prior to be of the form
\begin{equation} 
\pi(\M) = \exp\left[- \left(\frac{n_\mathrm{particles}}{a_1} + \frac{n_\mathrm{BRs}}{a_2} + \frac{n_\mathrm{SSMs}}{a_3}\right) \right] \,,
\label{eq:prior}
\end{equation}
where $n_\mathrm{particles}$ is the number of new particles present in the \protomodel,  
$n_\mathrm{BRs}$ is the number of non-trivial branching ratios%
\footnote{This means, for only one decay mode with 100\% BR,  $n_\mathrm{BRs}=0$.}, 
and $n_\mathrm{SSMs}$ the number of signal strength multipliers.%
\footnote{Note that particle-particle and particle-antiparticle production modes can have different signal strength multipliers.} 
In order to favor democratic decays, if two decay channels for a given particle have BRs differing by less than 5\%, they are counted only once. 
The parameters $a_1$, $a_2$, and $a_3$ are chosen to be 2, 4, and 8, respectively. This way, one particle with one non-trivial decay and two production modes is equivalent to one free parameter in the Akaike Information Criterion (AIC)~\cite{AIC}. 
For the SM, $n_\mathrm{particles}=n_\mathrm{BRs}=n_\mathrm{ssm}=0$ and the prior in Eq.~\eqref{eq:prior} satisfies the normalization set by Eq.~\eqref{eq:priorsm}. 
The test statistic thus roughly corresponds to a $\Delta \chi^2$ of the \protomodel with respect to the SM, with a penalty for the new degrees of freedom. 
Note that the prior is not normalized in the space of all \protomodels.

In Table~\ref{tab:priors} we illustrate how the prior affects the model test statistic \K for a few choices of parameters.
It should be noted that, with our choice of prior, \K ceases to have a clear,
probabilistic interpretation. \K can even become negative. We interpret such an
outcome as a preference of the SM over the BSM model.

\begin{table}
\centering
	\begin{tabular}{l|c|c|c|c}
		description & n$_{\mathrm{particles}}$ & n$_{\mathrm{BRs}}$ &
n$_\mathrm{SSMs}$ & $ \Delta \K$ \\
		\hline
		\hline
		one new particle, one non-trival BR, & & & & \\
		two production modes & 1 & 1 & 2 & $-2$ \\
		\hline
		two new particles, trivial BRs, & & & & \\
		five production modes & 2 & 0 & 5 & $-3.25$ \\
		\hline
		three new particles, non-trivial & & & & \\
		branchings and SSMs & 3 & 3 & 8 & $-6.5$ \\
	\end{tabular}
	\caption{Examples of \protomodel parameters and how the
corresponding prior affects the \protomodel test statistic. The last column displays 
the decrease in the test statistic: $\Delta \K = 2 \ln \pi(\BSM) $.}
	\label{tab:priors}
\end{table}

\subsection{Toy Walk}
\label{sec:toywalk}

In order to illustrate the walker algorithm, we first apply it to an extremely reduced version of the database with only three 13~TeV analyses: the ATLAS search for stops in the 1~lepton channel (ATLAS-SUSY-2016-16)~\cite{Aaboud:2017aeu}, the CMS search for hadronically decaying tops (CMS-SUS-16-050)~\cite{Sirunyan:2017pjw}, 
and the generic CMS search in the multi-jet plus \met\, channel 
(CMS-SUS-19-006)~\cite{Sirunyan:2019ctn}. 
All three present results for simplified models with 2 and 4 tops or
b-jets in the final state.%
\footnote{ATLAS-SUSY-2016-16 has EM results in addition to UL maps; the results from the two CMS analyses consist of expected and observed ULs.}

As discussed previously, the aim of the algorithm is to find \protomodels which can fit dispersed signals and at the same time satisfy the other constraints.
In this example, the potential dispersed signals appear in the ATLAS and CMS stop searches, where $\sim 1$--$2\,\sigma$ excesses have been observed in some signal regions. The third analysis (CMS-SUS-19-006), on the other hand, has seen an under-fluctuation in data and will play the role of the constraining result, or critic.
Indeed, since the two CMS searches included cannot be considered as uncorrelated (see Fig.~\ref{fig_matrix}), they are not allowed to be combined into a single likelihood. The ATLAS analysis, however, can be combined with either one of the CMS searches, and since the small excesses are found in ATLAS-SUSY-2016-16 and CMS-SUS-16-050, these will correspond to the best combination discussed at the beginning of Section~\ref{sec:walker}.

\begin{figure}[t!]
	\begin{center}
		\includegraphics[width=.95\textwidth]{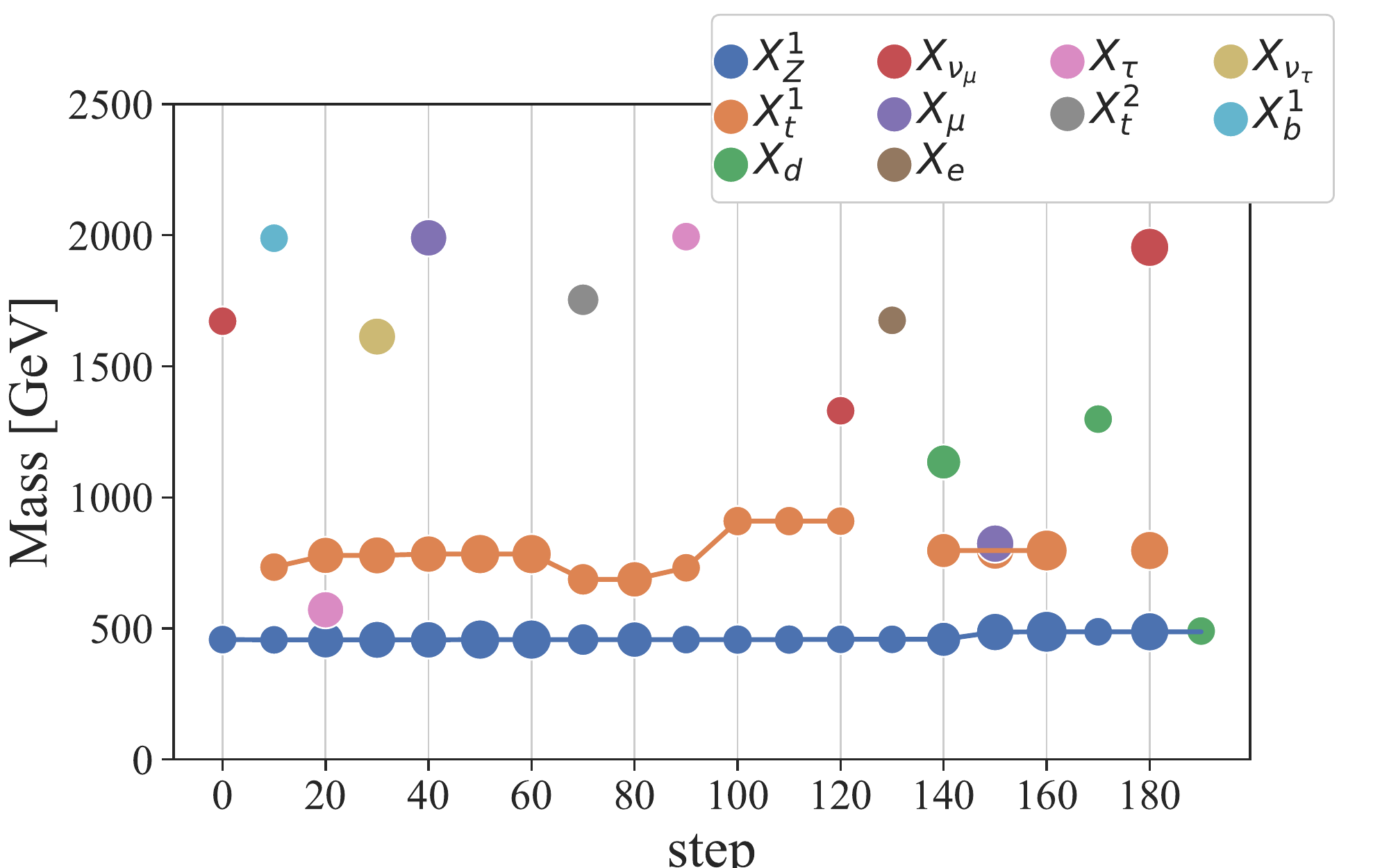}
		\caption{Evolution of the walker during 200 steps for the toy model of Secion~\ref{sec:toywalk}. Shown are the particle content and masses every 10 steps. The scaling of each point is
proportional to the \protomodel's test statistic \K at the corresponding step, so bigger points represent larger \K values.}
		\label{fig:toywalkA}
	\end{center}
\end{figure}

In Fig.~\ref{fig:toywalkA} we display the evolution of the walker
during 200 steps, by showing the masses of the particles present in the
\protomodel every 10 steps. Each point is scaled according to the \protomodel
test statistic \K, so bigger points correspond to higher \K values.
Since the database considered only contains results for simplified models with tops and bottoms in the final state, \protomodels without top and bottom partners find no corresponding data and no test statistic can be computed; hence they are set to zero.
Around step~20, a \protomodel with a top partner ($X_{t}^1$) is created, thus resulting in an increase in the test statistic.
The following steps modify the $X_Z^1+X_t^1$ model by randomly changing its parameters and adding or removing particles. Since the addition of new states only reduces the model test statistic due to the prior (see Eq.~\eqref{eq:prior}), the walker usually reverts back to the minimal scenario with one $X_Z$ and one $X_t$. We also see that for a few steps, such as step 70, the top partner is replaced by another state, resulting in a drastic reduction in \K. 
After 200 steps, the highest scoring \protomodel is that of step 94 and only contains the $X_Z^1$ and $X_t^1$ with masses of roughly 160~GeV and 1.1~TeV, respectively.

\begin{figure}[t!]
	\begin{center}
		\begin{minipage}{0.32\textwidth}
			\includegraphics[width=.9\textwidth]{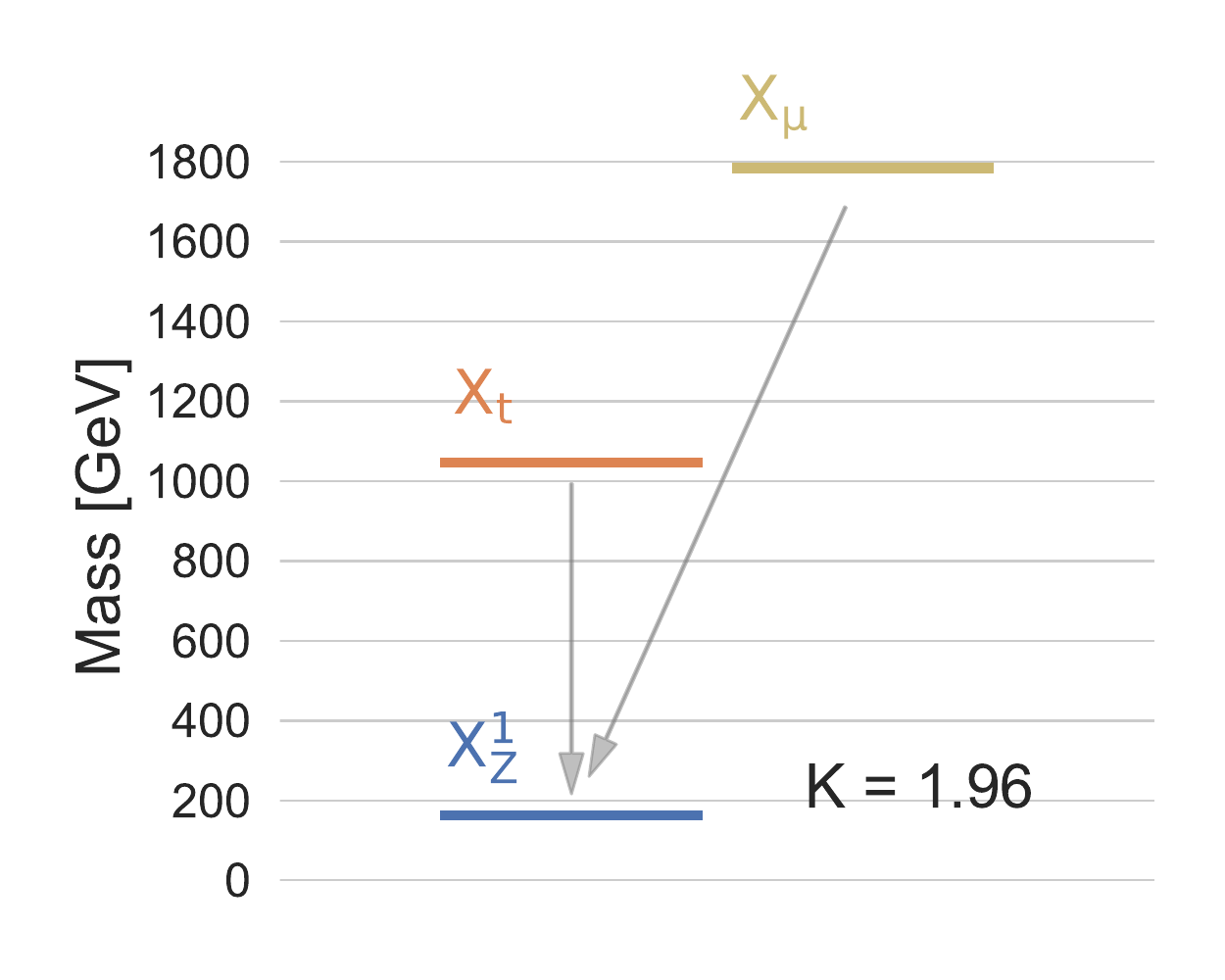}
		\end{minipage}
		\begin{minipage}{0.32\textwidth}
			\includegraphics[width=.9\textwidth]{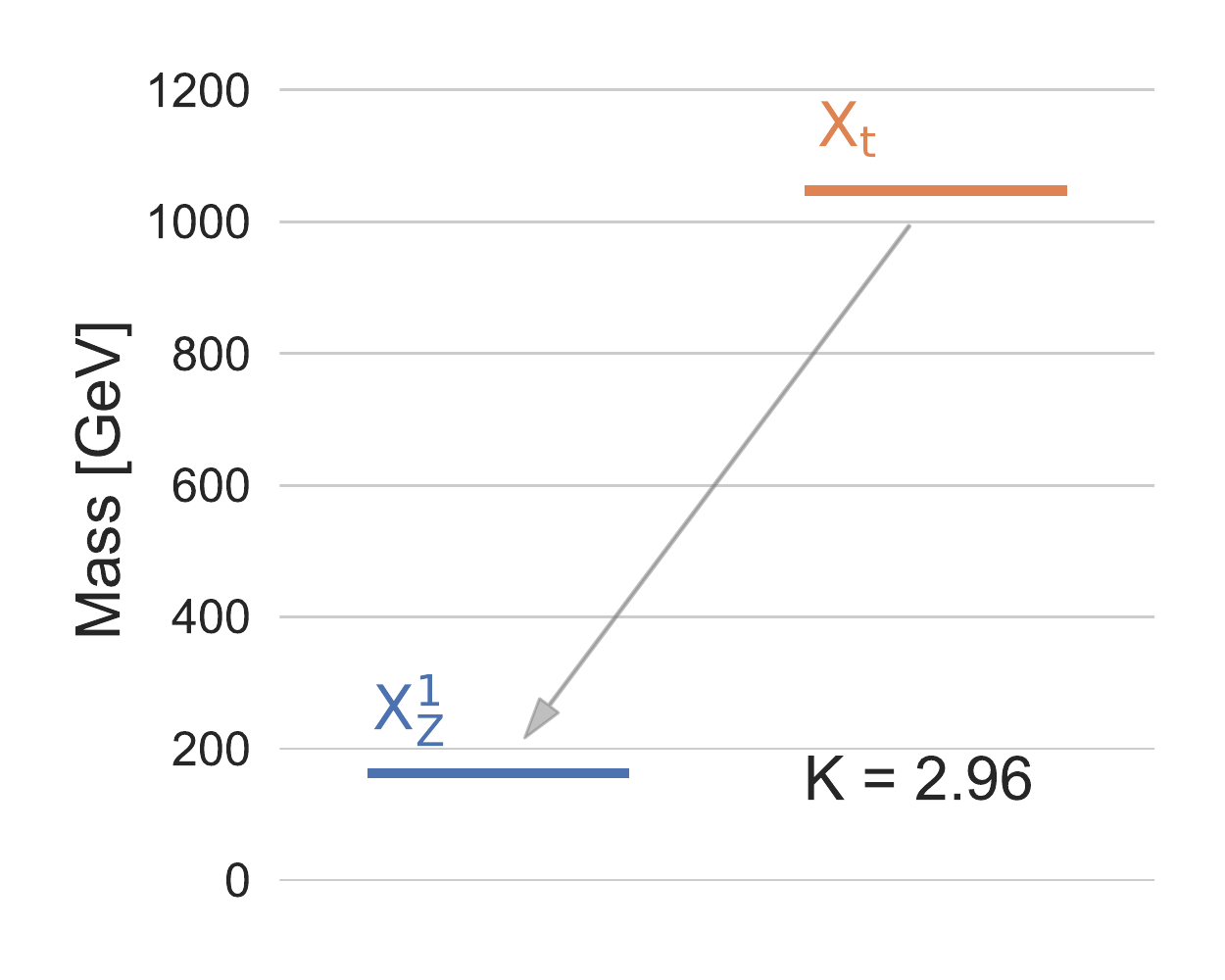}
		\end{minipage}
		\begin{minipage}{0.32\textwidth}
			\includegraphics[width=.9\textwidth]{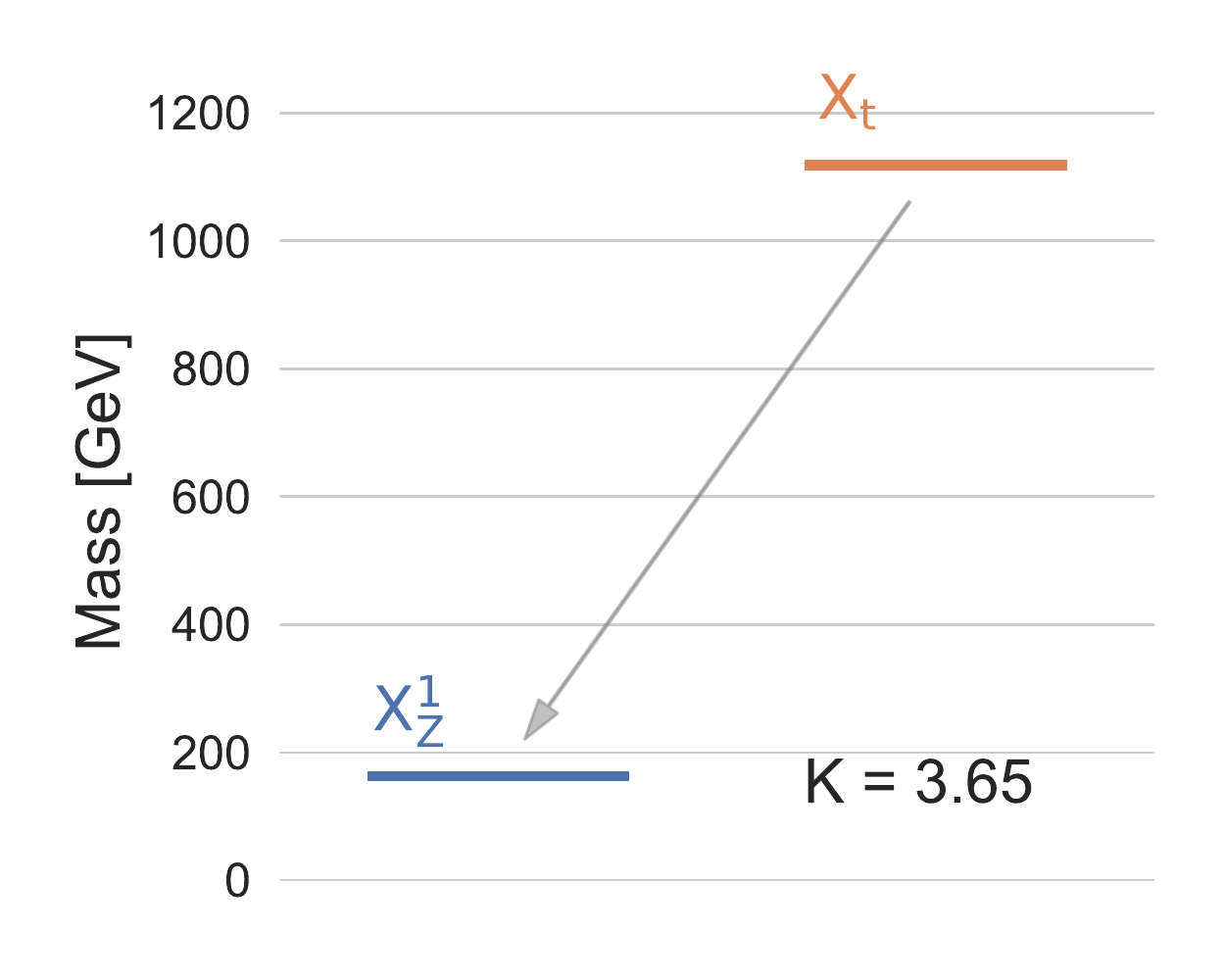}
		\end{minipage}\\
		\includegraphics[width=\textwidth]{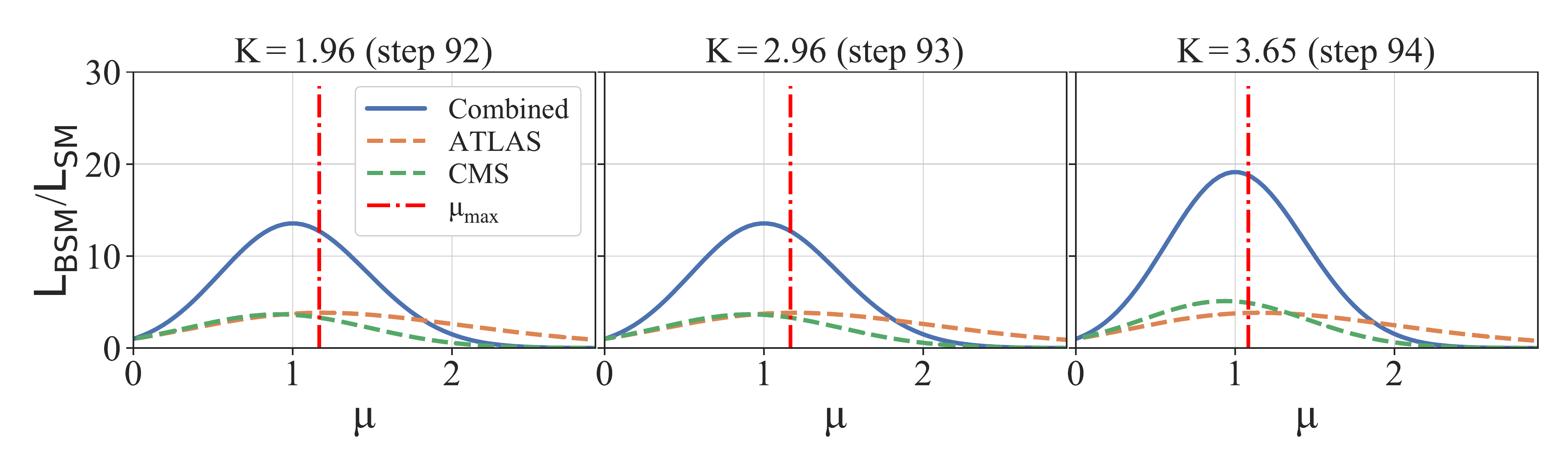}
		\caption{Examples of the likelihoods as a function of the signal strength $\mu$ at three steps for the toy walk discussed in the text. The blue solid curves represent the combined likelihood, while the dashed lines show the individual likelihoods for the ATLAS-SUSY-2016-16 and CMS-SUS-16-050 analyses. The upper limit on $\mu$ obtained from the CMS-SUS-19-006 analysis is shown by the dash-dotted red lines. The curves are normalized by the corresponding SM value (at $\mu = 0$). The upper panels display the corresponding \protomodel particle content and their decays.}
		\label{fig:toywalkB}
	\end{center}
\end{figure}

The behavior of the likelihoods at a few relevant steps is illustrated in Fig.~\ref{fig:toywalkB}, where the $\mathrm{L}_{\BSM}(\mu)$ curves,  normalized by $\mathrm{L}_{\SM}$, are shown for the best combination of analyses, ATLAS-SUSY-2016-16 and CMS-SUS-16-050. The maximum signal strength $\mu_{\rm max}$ allowed by the constraining analysis (CMS-SUS-19-006) is indicated by the dash-dotted vertical lines.
The first two panels show the steps immediately before the highest-score model. The first (leftmost) panel corresponds to a \protomodel containing  a $X_{\mu}$, $X_{t}^1$ and $X_Z^1$ with masses around 1.8~TeV, 1~TeV and 160~GeV, respectively. Since only the top partner contributes for fitting the excesses in the best combination, $X_\mu$ adds irrelevant degrees of freedom, thus reducing the \protomodel score due to the penalty from the prior. In the following step, shown in the center panel, $X_\mu$ is removed, without impact on the likelihood ratio, as expected. Nonetheless, due to the change in prior, the test statistic \K increases by about one unit. The next step, shown in the right panel, modifies the $X_t$ mass, which gives a better fit to the excess and an increase of the likelihood ratio. With $\K=3.65$ this turns out to be the highest-score model of the toy walk.

Although better results are likely to be found with a larger number of steps, this example illustrates how the algorithm walks in the \protomodel parameter space to find models consistent with dispersed signals and the constraints. In the following section, we will present more meaningful results, found using the same algorithm but for the full \smo\ database and with a much larger number of steps.

\section{Results using the full \smo Database}
\label{sec:results}

In this section we present the results of running the walker algorithm over the full \smo v1.2.4 database, which includes simplified model results from 40 ATLAS and 46 CMS analyses as detailed in Appendix~\ref{sec:anatables}. 

Interpreting the relevance of the \K values obtained for the \protomodels created by the algorithm requires knowledge about the \K distribution under the SM-only hypothesis. 
To this end, we first replace the observed data by values sampled from the expected SM backgrounds, resulting in an ensemble of ``fake databases'',   
which are then used to estimate the density of the test statistic \K under the SM hypothesis. This procedure is presented in Section~\ref{ssec:walkSM}. 
In Section~\ref{ssec:walkReal}, we then apply the walker to the real data. Using the \K distribution obtained under the SM hypothesis, we can evaluate the significance of the results and estimate a global $p$-value. 
Finally, in Section~\ref{ssec:walkSignal}, we will test the closure of our procedure by generating another ``fake'' database, where the signal of a given \protomodel has been injected in the data. This way we can verify the behavior of the algorithm and make sure that it is indeed capable of correctly reconstructing a signal present in the data. 

For all results presented in this section, we used $50$ parallel walkers, each with $1,000$ steps.
For reasons of CPU time consumption, we ran the code without making use of the correlations across signal regions provided for a few ATLAS and CMS analyses (cf.\ section~\ref{sec:likelihoods});
this is justified because the EM-type results, for which a covariance matrix or {\tt pyhf} JSON files are presently available, do not noticeably influence the results.\footnote{We checked this explicitly by running trials with signal region combination turned on.}
With this setup, each walker has a runtime of 12h to 24h on a single core.

\subsection{Walking over ``fake'' Standard Model Data}
\label{ssec:walkSM}

In order to compute the significance of the test statistic and,
ultimately, its global $p$-value, we produce ``MC toy'' versions of the
database, in which we replace the observed data by values sampled from the SM background models. More specifically, 
\begin{itemize}
	\item for EM-type results we sample a normal distribution
	with a mean of the background estimate, and a variance of the squared
	error of the background estimate. The sampled value is then entered as the lambda
	parameter of a Poissonian distribution. Finally, the value drawn from the Poissonian is used as the ``fake'' observation, i.e.\ the fake event yield;
	
	\item for UL-type results, for which expected upper limits are known to \smo, we estimate the SM uncertainty on the expected limit assuming a large number of observed events and small systematical uncertainties~\cite{Azatov:2012bz}. The observed upper limits are then sampled from a Gaussian distribution:
	\begin{equation}
	\sigma^\mathrm{UL}_\mathrm{obs} \sim \mathrm{Norm}\left(\sigma^\mathrm{UL}_\mathrm{exp}, \left(\delta \sigma^\mathrm{UL}_\mathrm{exp}\right)^2 \right),
	\end{equation}
	where $\delta \sigma^\mathrm{UL}_\mathrm{exp} =  \sigma^\mathrm{UL}_\mathrm{exp}/1.96$ is the estimated width of the distribution;

	\item UL-type results, for which only observed upper limits are
	available, enter the fake database as they are, since it is not possible to estimate the SM uncertainty in this case.
\end{itemize}

Using the sampled values described above, we produce 50 fake databases and apply the walker algorithm to all of them. The \protomodels with the highest \K values 
from each of these runs are then used to estimate the density of the test statistic \K under the SM hypothesis via a Kernel density estimator:

\begin{equation}
\label{eq:kde}
\rho(K) := \frac{1}{N w}\sum\limits_{i=1}^{N} \mathrm{kern}_w \left( \K-\K^i_\mathrm{fake}\right)
\end{equation}

\begin{figure}[t!]
\begin{center}
\includegraphics[width=.7\textwidth]{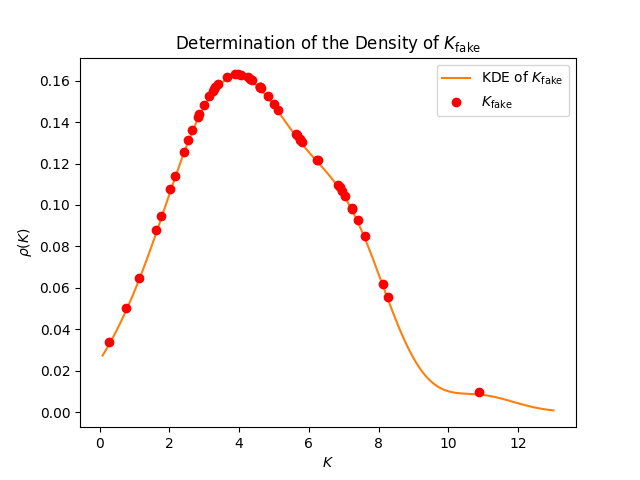}
\caption{The density of the test statistic \K under the SM-only hypothesis obtained using a Kernel Density Estimator (KDE). The \K values generated by the runs over 50 fake databases are shown as red points (see text for details).}
\label{fig_kde}
\end{center}
\end{figure}

Here, $N = 50$ is the number of runs with fake databases, and
$\K^i_\mathrm{fake}$ the test statistic for the $i$-th run; 
$\mathrm{kern}_w$ denotes the choice of Kernel in the Kernel density
estimation. We choose a Gaussian with a width $w$ determined by {\it Scott's rule}~\cite{scottsrule}. 
Figure~\ref{fig_kde} shows the $\K^i_\mathrm{fake}$ values generated for each run as well as the result obtained for $\rho(K)$. As we can see, under the SM hypothesis, 
values up to $\K \approx 11$ are expected,  
with the density peaking at $\K\approx 4$. 

\begin{figure}[t!]
	\begin{center}
		\includegraphics[width=.7\textwidth]{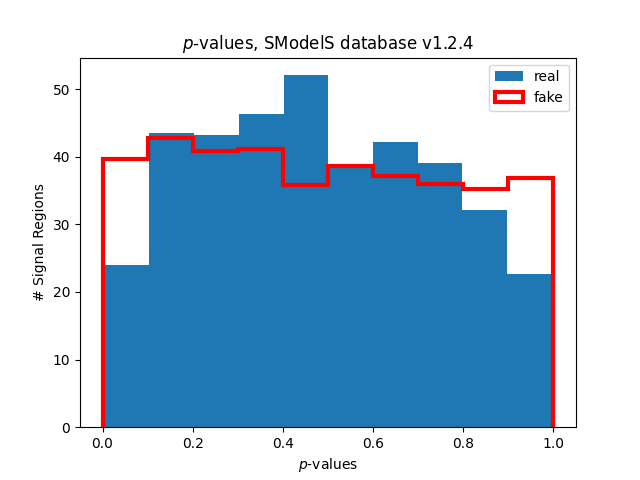}
		\caption{Distribution of $p$-values from EM results in \smo. The red histogram shows the distribution for a few ``fake'' databases
(but normalized to a single database), where the observed data has been
obtained by sampling the background expectations as explained in the text. The solid blue histogram
displays the same distribution, but for the real data. In order to avoid
distortions due to small numbers of events, only signal regions with at least 3.5 expected events were included.}
		\label{fig:biasses}
	\end{center}
\end{figure}

As discussed in Section~\ref{sec:walker}, the walker algorithm looks for dispersed signals, with \K computed as the maximum over all possible combinations of results. Therefore, large \K values are expected to be found even when considering the SM hypothesis, where upward background fluctuations can mimic dispersed signals.
In addition, it has been suggested~\cite{Nachman:2014hsa} that the experimental results
tend to be conservative and overestimate the background uncertainties. 

We can strengthen this claim by
considering the $p$-values computed for all signal regions
of the EM results contained in the database. If the observed data is well described by the estimated backgrounds and the uncertainties thereon, one would expect a flat distribution of $p$-values. In fact, for the fake databases generated under the SM hypothesis, this is the case, as shown by the red histogram in Fig.~\ref{fig:biasses}. When we consider the real data, however, the distribution is not flat, as seen from the blue histogram in the same figure. This is expected if the background uncertainties are overestimated. 
As a consequence, the density $\rho(K)$ of the SM hypothesis in Fig.~\ref{fig_kde} is conservative; we can safely assume that the actual density would be shifted towards lower values of \K  
and that our results generally lie on the conservative side.

\subsection{Walking over the LHC Data}
\label{ssec:walkReal}

\begin{figure}[t!]
	\begin{center}
		\includegraphics[width=0.9\textwidth]{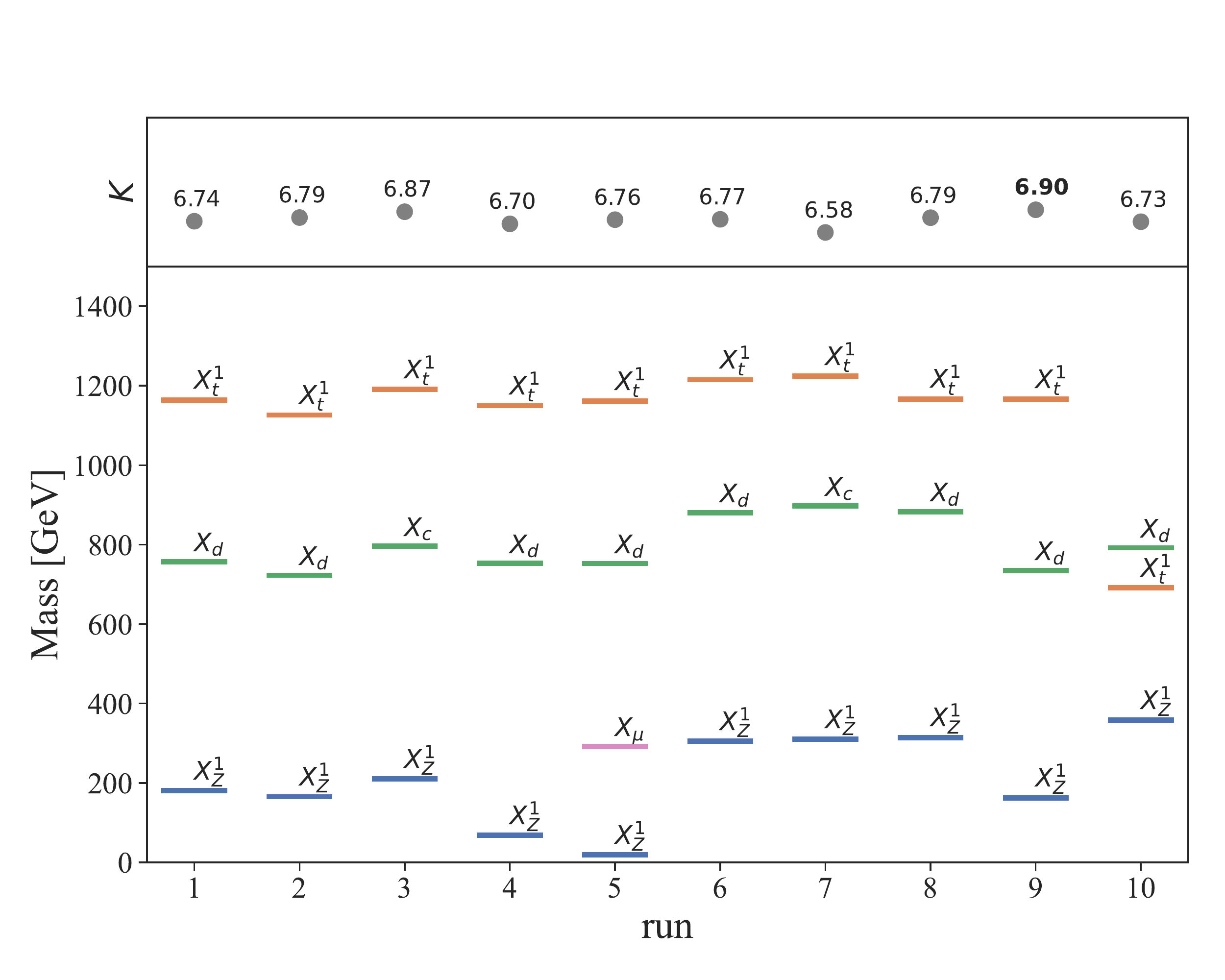}		
		\caption{Particle content and masses for the \protomodels with highest test-statistic \K obtained in each of the 10 runs performed over the real database. The corresponding \K values are shown at the top. Note that for all practical purposes $X_d$ and $X_c$ are indistinguishable.}
		\label{fig:winners}
	\end{center}
\end{figure}

We are now ready to apply the walker algorithm to the actual \smo database with the real LHC results. 
We perform 10 runs, each employing $50$ walkers and $1,000$ steps/walker. The results are summarized in Fig.~\ref{fig:winners}, where we display the  \protomodels with the highest \K value from each run. 
Besides the $X_Z^1$ LBP, 
all models include one top-partner, $X_t^1$, 
and one light-flavor quark partner, $X_{d,c}$, and their test statistics are at $\K = 6.76 \pm 0.08$ 
thus showing the stability of the algorithm.
The $X_\mu$ particle introduced in run~5 is
due to small $\approx 1 \sigma$ excesses in the CMS search for sleptons, CMS-SUS-17-009~\cite{Sirunyan:2018nwe} and the ATLAS search for electroweakinos, ATLAS-SUSY-2016-24~\cite{Aaboud:2018jiw};  
the prior's penalty for introducing the additional particle, $\Delta K = - 1.25$, is overruled by the increase in the likelihood ratio, $\Delta K = 1.94$.

Although the scenarios from Fig.~\ref{fig:winners} are statistically all equally plausible, it is instructive to discuss the absolute ``winning'' one, 
i.e.\ the \protomodel with $\K = 6.90$ generated in step 582 of the 29th walker in run \#9, in some more detail.
It has $X_t^1$, $X_d$ and $X^1_Z$ masses of 1166,  735 and 163~GeV, respectively, and   
produces signals in the $t\bar t+E_T^{\rm miss}$ and jets$+E_T^{\rm miss}$ final states with SUSY-like cross sections.
Concretely, the effective signal strength multipliers are $\hat{\mu} \times \kappa_{X_t^1 \bar X_t^1}\approx 1.2$ and $\hat{\mu} \times  \kappa_{X_d \bar X_d}\approx 0.5$, corresponding to $\sigma(pp\to X_t^1 \bar X_t^1)\approx 2.6$~fb  and $\sigma(pp\to X_d \bar X_d)\approx 24$~fb at $\sqrt{s}=13$~TeV, and both the $X_t^1$ and $X_d$ directly decay to the lightest state ($X^1_Z$) with 100\% BR.

\begin{table}[t!]
\hspace*{-2mm}%
\begin{tabular}{l|c|r|r|c|r|r}\small
\bf{Analysis} & \bf{Dataset} & \bf{Obs}\;\phantom{.} & \bf{Exp}\;\,\phantom{.} & \bf{Z} & \bf{P} & \bf{Signal} \\
\hline
ATL multijet, 8 TeV~\cite{Aad:2014wea} & SR6jtp & 6 & 4.9 $\pm$ 1.6 & 0.4 $\sigma$ & $X_{d}$ & 0.25 \\ 
ATL multijet, 13 TeV~\cite{Aaboud:2017vwy} & 2j\_Me ... & 611 & 526 $\pm$ 31 & 2.2 $\sigma$ & $X_{d}$ & 44.18 \\ 
ATL 1$\ell$ stop, 13 TeV~\cite{Aaboud:2017aeu} & tN\_high & 8 & 3.8 $\pm$ 1 & 1.9 $\sigma$ & $X_{t}$ & 3.93 \\ \hline
CMS multijet, 8 TeV~\cite{Chatrchyan:2014lfa} &  & 30.8 fb & 19.6 fb & 1.1 $\sigma$ & $X_{d}$ & 2.66 fb \\ 
CMS $0\ell$ stop, 13 TeV~\cite{Sirunyan:2017pjw} &  & 4.5 fb & 2.5 fb & 1.6 $\sigma$ & $X_{t}$ & 2.62 fb \\ 
\end{tabular}
	\caption{Analyses contributing to the \K value of the highest score \protomodel from Fig.~\ref{fig:winners}. Listed are the signal region (SR) where applicable, the number of observed events or observed upper limit (Obs), the expected background and its uncertainty or the expected upper limit (Exp), the significance (Z) of the excess, and which \protomodel particle (P) is introduced for fitting the observed data.  The last column (Signal) shows the predicted  contribution of the \protomodel, in number of events for EM results (ATLAS, ATL for short) and in fb for UL results (CMS).}
		\label{tab:rawnumbers}
\end{table}

The analyses, which contain possible dispersed signals and drive this model (as well as the other ones in Fig.~\ref{fig:winners}), are listed in Table~\ref{tab:rawnumbers}. 
The light quark partner with mass around 700~GeV is introduced in order to fit the observed excesses in the multijet (i.e., $0\ell$ + jets) +\met\xspace analyses from ATLAS at $\sqrt{s}= 8$ and  13~TeV~\cite{Aad:2014wea,Aaboud:2017vwy} and from CMS at $\sqrt{s}= 8$~TeV~\cite{Chatrchyan:2014lfa}. The \K value obtained is, however, mostly driven by the $\approx 1.5\,\sigma$ and $2\,\sigma$ excesses observed in the 13 TeV CMS and ATLAS stop searches~\cite{Sirunyan:2017pjw,Aaboud:2017aeu}, which lead to the introduction of a top partner ($X_t^1$) with a mass around 1.2~TeV. 
Despite corresponding to small excesses, identifying the presence of such potential dispersed signals is one of the main goals of the algorithm presented here. Furthermore, the fact that these excesses appear in distinct
ATLAS and CMS analyses and 
can be explained by the introduction of a single top partner 
is another interesting outcome of the whole procedure.

\begin{table}
\begin{tabular}{l|c|c|c|c|c}
\bf{Analysis (all CMS 13~TeV)}& \bf{Prod} & $\sigma_{XX}$ (fb) & $\sigma^\mathrm{UL}_\mathrm{obs}$ (fb) & $\sigma^\mathrm{UL}_\mathrm{exp}$ (fb) & $r_\mathrm{obs}$ \\
\hline
CMS multijet, $M_{H_T}$, 137~fb$^{-1}$~\cite{Sirunyan:2019ctn} & ($\bar{X}_{d}, X_{d}$) & 23.96 & 18.45 & 21.57 & 1.30\\
CMS multijet, $M_{H_T}$, 137~fb$^{-1}$~\cite{Sirunyan:2019ctn} & ($\bar{X}_{t}, X_{t}$) & 2.62 & 2.04 & 2.08 & 1.28\\
CMS multijet, $M_{H_T}$, 36~fb$^{-1}$~\cite{Sirunyan:2017cwe} & ($\bar{X}_{d}, X_{d}$) & 23.96 & 19.26 & 28.31 & 1.24\\
CMS multijet, $M_{\rm T2}$, 36~fb$^{-1}$~\cite{Sirunyan:2017kqq} & ($\bar{X}_{d}, X_{d}$) & 23.96 & 26.02 & 31.79 & 0.92\\
CMS $1\ell$ stop, 36~fb$^{-1}$~\cite{Sirunyan:2017xse} & ($\bar{X}_{t}, X_{t}$) & 2.62 & 2.91 & 4.44 & 0.90\\
\end{tabular}
	\caption{List of the most constraining results for the highest score \protomodel. The second column displays the constrained production mode, while the third column shows the respective cross section value. We also show the corresponding observed and expected upper limits and the ratio $r_\mathrm{obs} = \sigma_{XX}/\sigma^\mathrm{UL}_\mathrm{obs}$.	} \label{tab:rvalues}
\end{table}

It is worth noting that the signal injected by the winning \protomodel is typically smaller than the one favored by the excesses, as seen when comparing the {\bf Obs} and the {\bf Signal} columns in 
Table~\ref{tab:rawnumbers}.
This is due to a tension with the constraints imposed  on the signal strength by all the other analyses in the database (the {\it critic}). The model discussed above is mostly constrained by the 13 TeV CMS multijet analyses for 36~fb$^{-1}$~\cite{Sirunyan:2017cwe,Sirunyan:2017kqq} and 137~fb$^{-1}$~\cite{Sirunyan:2019ctn}, which all observed under-fluctuations with respect to the expected background, see Table~\ref{tab:rvalues}. 
The most serious challenges comes from the 2~jets+\met\ and $t\bar t$+\met\ limits 
of \cite{Sirunyan:2019ctn}, for which the \protomodel under consideration exceeds both the observed and expected UL. 
Concretely, the analysis \cite{Sirunyan:2019ctn} gives $r_{\rm obs}$ $(r_{\rm exp})=1.3$ ($1.11$) for the $X_d$, and $r_{\rm obs}$ $(r_{\rm exp})=1.28$ ($1.26$) for the $X_t^1$. 
\footnote{Recall that we accept mild transgressions up to $r\lesssim 1.3$ to account for the fact that when simultaneously checking limits from a large number of analyses, a few are statistically allowed to be violated.}

Finally, we point out that very similar conclusions are drawn for all the high score \protomodels shown in Fig.~\ref{fig:winners}. 
Of course, all these conclusions 
may be challenged (or strengthened) by new results for full Run~2 luminosity, which are not yet included in the \smo database, e.g.\ the stop searches~\cite{Aad:2020sgw,Sirunyan:2020tyy}.

\begin{figure}[t!]
\begin{center}
\includegraphics[width=0.7\textwidth]{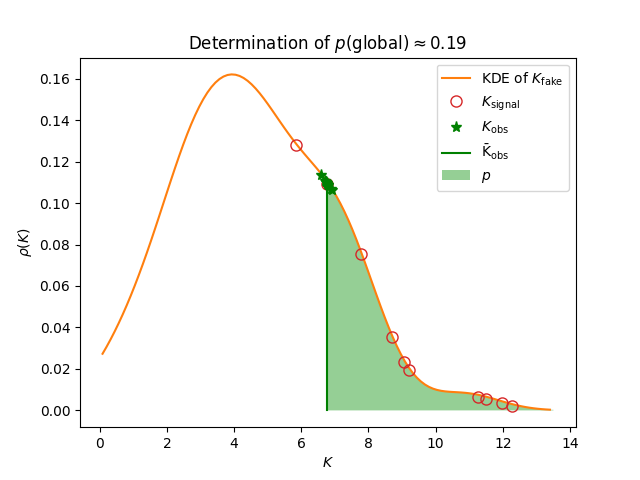}
\caption{Values of the test statistic \K obtained for the walks discussed in the text.  The green stars show the results for the 10 runs over the real database, while the red open circles show the values obtained under the ``BSM hypothesis'' (see discussion in Section~\ref{ssec:walkSignal}). The solid orange curve shows the distribution obtained under the SM-only hypothesis (see Fig.~\ref{fig_kde}). The shaded green area corresponds to the global $p$-value for average \K from the runs over the real database.}
\label{fig_summarywalks}
\end{center}
\end{figure}

Some comments are in order on the significance of the excesses and \K values found for the models  in Fig.~\ref{fig:winners}. In Fig.~\ref{fig_summarywalks} we show the \K distribution found under the SM-only hypothesis (see the discussion in Section~\ref{ssec:walkSM}) and the values for the 10  highest-scoring \protomodels obtained with the ``real''  database (green stars).
As we can see, despite being at the tail of the distribution, the values still seem compatible with the SM-only hypothesis. In order to quantify this compatibility, we compute the $p$-value given by the relative frequency of the \K values under the SM hypothesis which were found above the observed \K values ($K_{\mathrm{obs}}$):
\begin{equation}
p_\mathrm{global} := \lim\limits_{N\rightarrow\infty} \frac{1}{N} \sum\limits_{i=1}^{N}
\mathds{1}_{[\bar{\K}_\mathrm{obs},\infty)}(\K^i_\mathrm{fake}) \approx
\int\limits_{\bar{\K}_\mathrm{obs}}^{\infty} d\K \rho(K) \,.
\end{equation}

Here $N=50$ is the number of fake SM-only runs, $\mathds{1}_{X}(x)$ denotes the
indicator function that equals unity for all $x \in X$ and zero otherwise,  
$\bar{\K}_\mathrm{obs}$ is the
average of $\K_\mathrm{obs}$ of the ten runs over the ``real'' database, 
and $\rho(K)$ refers to the estimated density of the test statistic as
defined in Eq.~\eqref{eq:kde}. The resulting value corresponds to the shaded area under the curve shown in Fig.~\ref{fig_summarywalks}. Using the $K$ values in Fig.~\ref{fig:winners} we obtain the global $p$-value 
for the SM hypothesis:
\begin{equation}
   p_\mathrm{global} \approx 0.19 \,.
\end{equation}
Our results thus indicate a very mild disagreement with the SM hypothesis. As already pointed out in Section~\ref{ssec:walkSM}, we expect this to be conservative due to potentially overestimated background uncertainties. Furthermore, this result requires no look-elsewhere correction, since the SM density $\rho(K)$ was derived through the exact same procedure as applied for the real database. 

Complementary information on the distributions of the test statistic \K and posterior densities can be found in appendix~\ref{sec:distributions}.

\subsection{Walking over ``fake'' Signals}
\label{ssec:walkSignal}

As discussed above, the \K values obtained for the real database are in a mild disagreement with the SM hypothesis. It is relevant, however, to investigate whether the procedure proposed in this work would be able to reproduce the underlying model. 
For this kind of closure test, we employ a scheme analogous to the one used to produce fake databases under the SM-only hypothesis in Section~\ref{ssec:walkSM}. This time, however, using the highest-score model from Fig.~\ref{fig:winners}, we create fake observed data under the ``BSM hypothesis''.
Concretely, we sample from the signal plus background model, assuming that the corresponding uncertainties are dominated by the errors on the background. However, when producing the fake ``BSM hypothesis'' databases, we scale the background uncertainties by a ``fudge factor'' of 0.65, aiming to produce fake data closely resembling those observed in the real database. This is motivated by the fact that, as discussed in Section~\ref{ssec:walkSM}, the observed data in the real database point to an overestimation of the background uncertainties.

The theory predictions from the injected \protomodel are included as follows:
for EM-type results the signal + background model has central values given by the SM expected value plus $\sigma \times \mathrm{BR} \times \epsilon$ for the injected signal. The number of observed events is then determined by sampling the signal + background model.
For UL-type results, on the other hand, we simply shift the observed limit by the predicted cross section in the region of the UL map that has masses within 100~GeV of the assumed \protomodel values. This procedure is a good approximation within the Gaussian limit if the signal efficiencies do not change significantly within the shifted region.

\begin{figure}[t!]\centering
\includegraphics[width=0.94\textwidth]{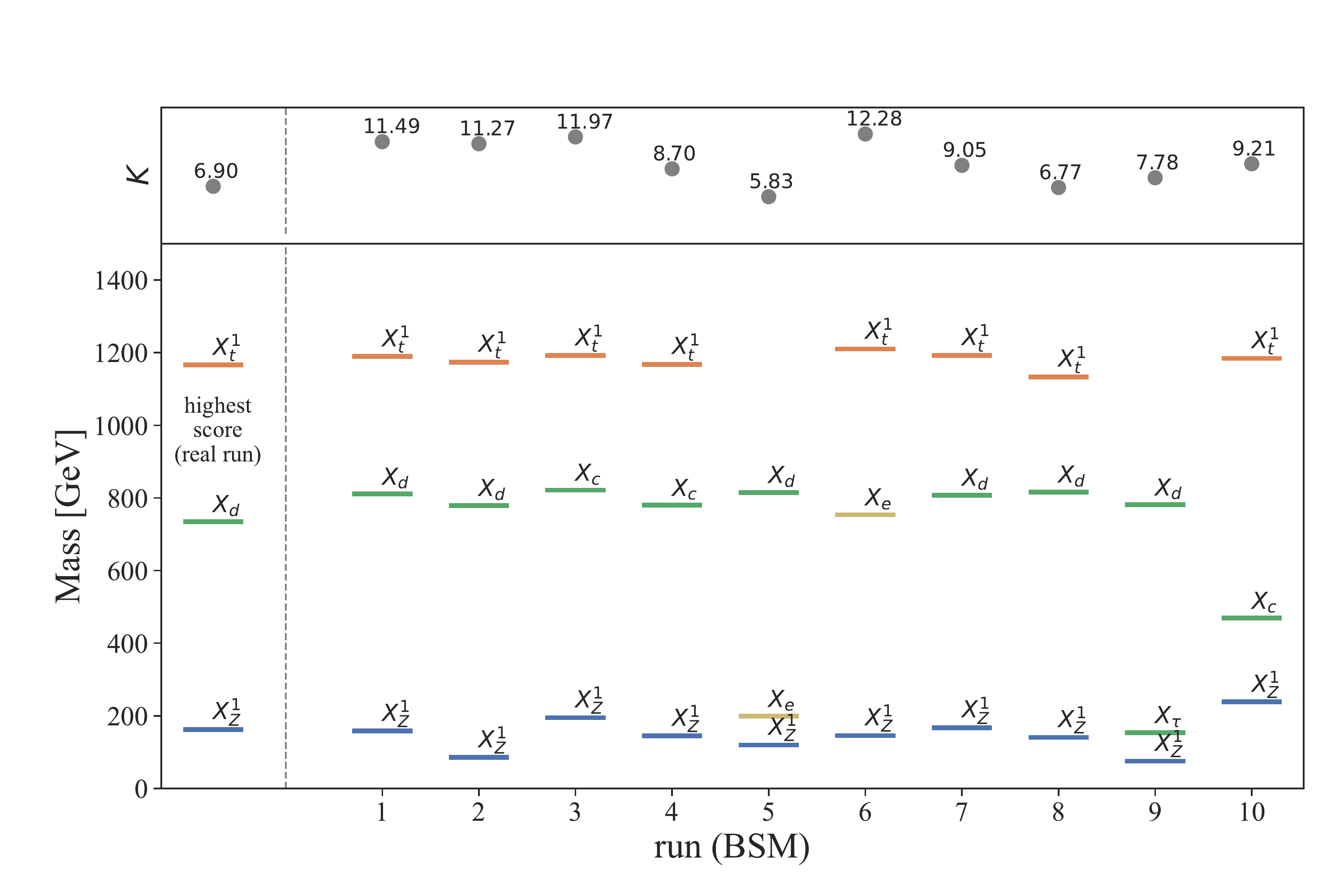}
\caption{Particle content and masses for the \protomodels with highest test-statistic \K obtained in each of the 10 runs performed over the fake databases generated under the ``BSM hypothesis''. The corresponding $K$ values are shown at the top. The spectrum of the \protomodel used for the injected signal is shown on the left.}
\label{fig_compfakesignals}
\end{figure}

We generate 10 such ``fake signal'' databases and perform a run with each of them. 
The resulting distribution of the test statistics \K is shown in Fig.~\ref{fig_summarywalks}.
As can be seen, under the BSM hypothesis (the fake signals) 
\K varies between $\K \approx 5.8$ and $\K \approx 12.3$. 
Figure~\ref{fig_compfakesignals} shows the particle spectra of the high-score models of these fake signal runs and compares them to the injected signal. 
We see that $X_{t}^1$ 
was reconstructed nicely at around the right mass scale, in 8 out of 10 runs. The quark partner,
$X_{q}$, was found in 9 out of 10 runs, though
once at a significantly lower mass. Finally, runs~5, 6, and 9 introduced spurious lepton partners, all due to various background fluctuations created in the statistical sampling.

\begin{figure}[t!]
\begin{center}
\includegraphics[width=0.94\textwidth]{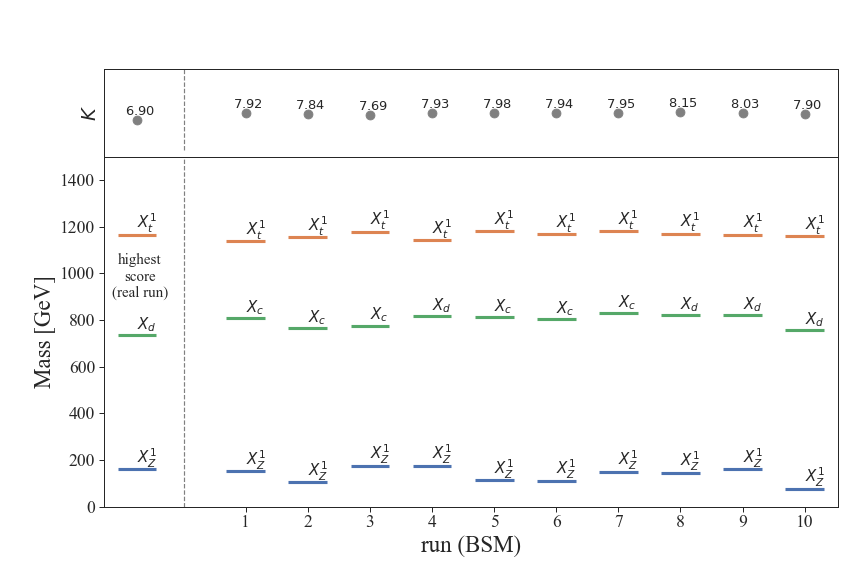}
\caption{``Technical'' closure test, replacing the observed data with the expected background plus signal yields (without sampling).
The spectrum of the \protomodel used for the injected signal is shown on the left.}
\label{fig_technicaltest}
\end{center}
\end{figure}

All in all, the differences between the injected and reconstructed signals shown in Fig.~\ref{fig_compfakesignals} are statistical in nature and in fact to be expected in our procedure, given that the injected signal is of the size of the background uncertainties. To verify this, we have also performed a purely technical closure test, where background and signal sampling were turned off, so the number of observed events for EM results is simply set to the corresponding central values of the signal plus background model. In this case, the particle content of the injected signal  is correctly reproduced in all runs, including the masses. 
The result from injecting the high-score model from Fig.~\ref{fig:winners}
is shown in Fig.~\ref{fig_technicaltest}.
It nicely confirms {\it (i)}~the robustness of our procedure, and {\it (ii)}~the conclusion that the small deviations in Fig.~\ref{fig_compfakesignals} are due to fluctuations generated by the statistical sampling.

\section{Conclusions and Outlook}
\label{sec:conclusions}

In view of the null results (so far) in the numerous  channel-by-channel searches for new particles, it becomes increasingly relevant to change perspective and attempt a more global approach to find out where BSM physics may hide. To this end we presented in this paper a novel statistical learning algorithm, that is capable of identifying potential dispersed signals in the slew of published LHC analyses. The task of the algorithm is to build candidate \protomodels from small excesses in the data, while at the same time remaining consistent with all other constraints.

At present, this so-called {\it \protomodel builder} is based on the concept of simplified models, exploiting the \smo software framework and its large database of simplified-model results from ATLAS and CMS searches for new physics. It employs a random walk through the \protomodel parameter space, adding or removing new particles, and randomly changing their cross sections and branching ratios. This is coupled to a combinatorial algorithm, which identifies the set of analyses and signal regions that maximally violates the SM hypothesis (from a total of 86 ATLAS and CMS search results). 

Running over the \smo v1.2.4 database, the \protomodel builder identified small excesses at  the level of $\approx 2\sigma$ in ATLAS and CMS stop searches, as well as in multijet\,+\,\met\, searches.  These may be explained by introducing SUSY-like top and quark partners $X_t^1$ and $X_q$ at a mass scale of roughly 1~TeV, 
together with a neutralino-like lightest particle $X_Z^1$ below about 400~GeV, while staying in agreement with constraints from other searches.
The highest-score \protomodel has a test statistic of $\K=6.9$, while the global $p$-value for the SM hypothesis is $p_{\rm global}\approx 0.19$.
The test statistic is effectively $K \approx \Delta\chi^2 + 2\log \pi(\M)$, where $\Delta\chi^2$ is a measure of the violation of the SM hypothesis, and $\pi(\M)$ penalizes for new degrees of freedom introduced in the protomodel. 

We stress that, while interesting {\it per se}, these results are intended mostly as a proof-of-principle. Indeed the current realisation of our 
\protomodel builder is
still limited by the variety and type of simplified-model results available. 
First, the mass planes of simplified model results need to extend far enough to high and low masses to allow a good coverage of all types of new particles and coupling strengths.\footnote{For example, fermionic partners have larger production cross sections than scalar ones; simplified model results developed for the scalar hypothesis (squarks, stops, etc.) therefore often do not extend to high enough masses to fully cover fermionic partners.} Wider mass ranges than is current practice would be very useful in this respect.
Second, and perhaps more importantly, only EM-type results (i.e.\ $A\times\epsilon$ maps) allow for the computation of a proper likelihood. Analyses for which only 95\%~CL limits on the signal cross sections are available force us to make crude approximations, which can significantly distort the statistical evaluation. So far, however, merely one third of the available simplified model results are EMs. 
Third, as also stressed in \cite{Abdallah:2020pec}, correlation data enabling the combination of signal regions are essential for avoiding either overly conservative or over-enthusiastic interpretations, and for stabilising global fits.\footnote{In this respect we also refer to Ref.~\cite{Athron:2018vxy}, which  found that the use of single best-expected signal regions was numerically unstable as well as statistically suboptimal. Furthermore, any approach forced to conservatively use single best-expected signal regions invalidates the interpretation of the profile log-likelihood-ratio via Wilks' Theorem, necessitating the uptake of approximate methods.} So far, however, appropriate correlation data in combination with EMs are available for only four analyses. 

The first and second issues above can in principle be resolved to some extent by ourselves, through the development of ``home-grown'' EMs by means of simulation-based recasting. This comes however with the caveat that ``home-grown'' EMs will always be less precise than ``official'' ones from the experimental collaborations. Moreover, information on background correlations can only be provided by the experiments. We therefore strongly encourage ATLAS and CMS to systematically provide EM results together with correlation data, 
following the recommendations in \cite{Abdallah:2020pec}.
The {\tt pyhf} JSON likelihoods, as provided for some ATLAS analyses, are particularly useful: in addition to communicating the full statistical model, which allows for a much more accurate evaluation of the likelihood, they may also allow for a better estimate of cross-analysis correlations than currently possible.

Also on the technical side our current procedure, while good enough for a proof-of-principle, can be improved in several aspects. For example, in the current version an Akaike-type information criterion has been used to judge the quality of a \protomodel. It will be interesting and relevant to try other information criteria, such as the Bayesian information criterion, or the deviance information criterion, systematically comparing their performances~\cite{Vrieze:2012}. 
Furthermore, we plan to work towards making the global, combined likelihood differentiable. This will allow for 
using gradient-based methods, which will be a major performance boost.
Note here that {\tt pyhf} also aims at full differentiability, which is another argument in favor of communicating full statistical models via this scheme.
In this respect it will also be interesting to machine learn the \smo\ database, e.g.\ via neural networks or symbolic regression, 
as these methods automatically come with a gradient.

Regarding the statistical interpretation, future work will concern, for instance, a more quantitative assessment of the fitted \protomodels, including estimations of posterior distributions for the relevant parameters.  
Information from LHC measurements may be added to the game in a Contur-like approach~\cite{Butterworth:2016sqg}. 
Evidently, the project will also profit from all developments of \smo itself, since any generalization of the \smo formalism allows for an ever larger space of \protomodels. 
Last but not least, going beyond the concept of \protomodels, it will be highly interesting to investigate how the high score models generated by our algorithm can be mapped to effective BSM field theories or even to full UV complete theories. 
All in all, there is much exciting work to be done.

\paragraph{Code and data management:} while not a published tool so far, the code of the \protomodel builder is publicly available at 
\url{https://github.com/SModelS/protomodels}
together with the detailed data files of the results presented in this paper.

\subsection*{Acknowledgements}

The computational results presented were obtained using the CLIP cluster~\cite{clip}.
We thank Nishita Desai, Sylvain Fichet, Jan Heisig, Suchita Kulkarni, Harrison Prosper and Helga Wagner for discussions related to this work. 
The work of S.K.\ was supported in part by the IN2P3 master project ``Th\'eorie -- BSMGA''.

\appendix
\section{Appendix}
\label{appendix}

\subsection{Building Proto-Models}
\label{sec:builder}

As discussed in Section~\ref{sec:protomodels}, \protomodels are defined by their particle content, masses, decay modes and signal strength multipliers. In order to perform
a MCMC-type walk over this parameter space, in each step of the walk the following random changes are made:
\begin{itemize}
	\item{{\bf Add a new particle:} one of the BSM particles listed in Table~\ref{tab:decays} not yet present in the model can be randomly added. Once added, the mass of the new particle is drawn from a uniform distribution between
		the LBP mass and 2.4 TeV. The new particle is initialized with random branching ratios (for the corresponding decays listed in Table~\ref{tab:decays})
		and signal strength multipliers set to one. Adding a particle is programmed to occur more often for models with low test statistics and/or with a small number of particles.}
	
	\item{{\bf Remove an existing particle:} one particle present in the model is randomly selected and removed. All the production cross sections and decays involving the removed particle are deleted and the remaining branching ratios are normalized, so they add up to 1. Removing a particle is set to occur more often for models with low test statistics and/or with a large number of particles.}
	
	\item{{\bf Change the mass of an existing particle:} the mass of a randomly chosen particle is changed by an amount $\delta_m$ according to a uniform distribution whose exact interval depends on the test statistic and number of unfrozen particles in the model, with better performing models making smaller changes. This change is always performed if no other changes have been made in the \protomodel in a given step.}
	
	\item{{\bf Change the branching ratios:} the branching ratio of a randomly chosen particle is changed. This change can occur in three distinct ways: i) a random decay channel can have its BR set to 1 and all other channels are closed, ii) a random decay channel can be closed and iii) each decay channel can have its BR modified by a distinct random amount $\delta_{BR}$ drawn from a uniform distribution between $-a$ and $a$, where $a = 0.1/\mbox{(number of open channels)}$. After any of these changes, the branching ratios are normalized to make sure they add up to unity.}
	
	\item{{\bf Change the signal strength multipliers:} the signal strength multiplier (SSM) can be randomly changed in three ways: i) a specific production cross section for a randomly selected final state can have its SSM re-scaled by a value drawn from a Gaussian distribution centered around 1 and with width 0.1, ii) a random process can have its SSM set to zero, one, or to the SSM of another process and iii) all the processes involving a randomly chosen particle can have their SSMs re-scaled by a random number between 0.8 and 1.2.}
	
\end{itemize}

In addition to the above changes, we also include the following changes in the \protomodel at each step of the MCMC walk:

\begin{itemize}
    \item{{\bf Check for particles below the mass wall or within the mass corridor}: any changes that would violate
    the conditions on \protomodels outlined at the end of Section~\ref{ssec:particlecontent} are reverted.}
	\item{{\bf Remove redundant particles}: particles which do not contribute to a signal entering the likelihood for the best combination of results are removed.}
	
	\item{{\bf Merge degenerate particles}: particles which have a mass difference smaller than 200 GeV are merged. Once the particles are merged, their signal strength multipliers are added and the relevant branching ratios are re-scaled. The merging is only accepted if the resulting \protomodel has a test statistic equal or higher than the original model.}

\end{itemize}

Note that the varying dimensionality of the statistical model caused by the non-constant number of particles, decay channels, and production modes is a feature shared with the ``reversible-jump MCMC'' algorithm~\cite{rjmcmc}.\footnote{We thank Andrew Fowlie for pointing this out.}

\subsection{Distributions and Posteriors}
\label{sec:distributions}

\subsubsection*{Distributions of \K}

\begin{figure}[t!] \begin{center}
	\hspace*{-6mm}\includegraphics[width=.54\textwidth]{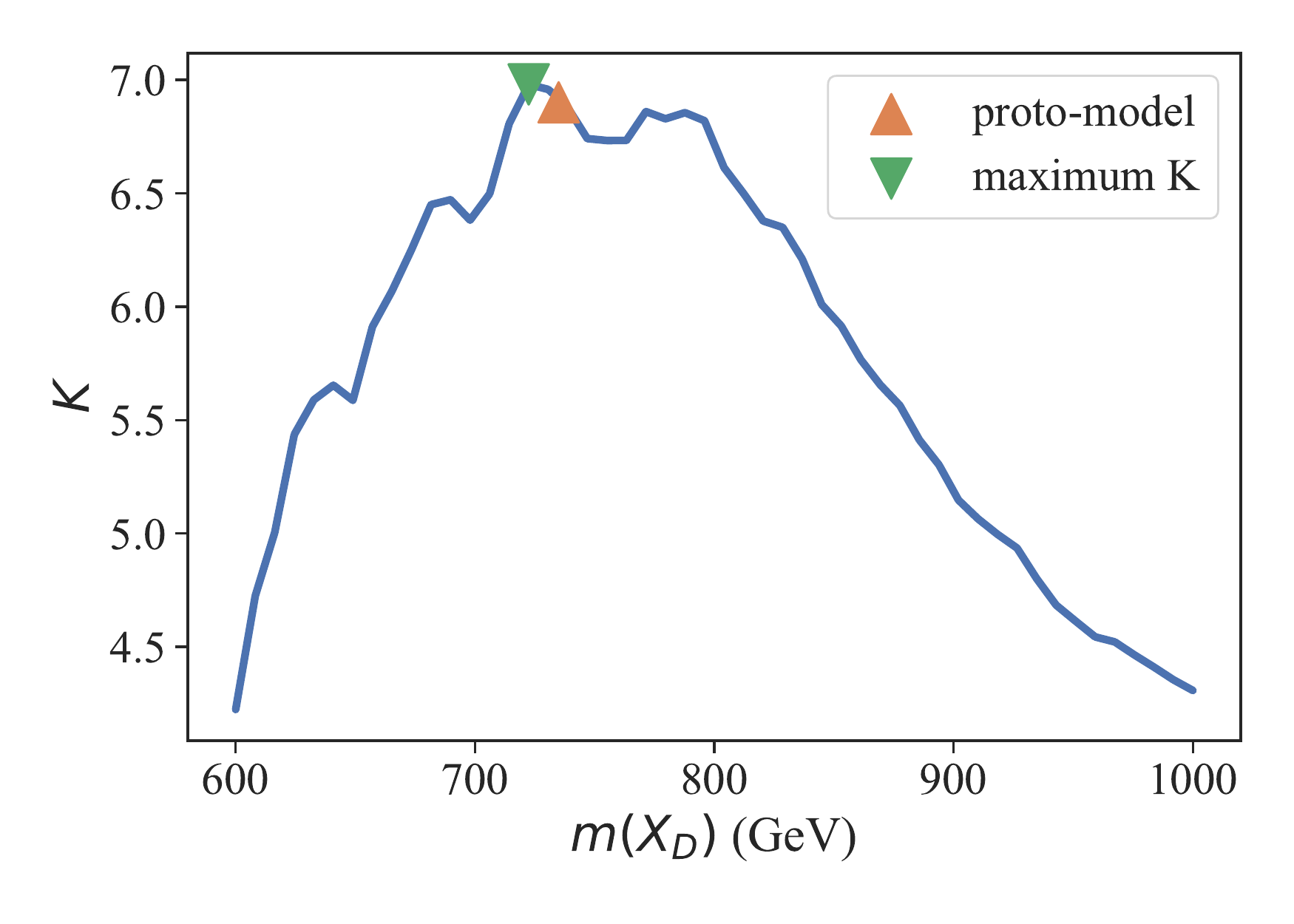}%
	\hspace*{-2mm}\includegraphics[width=.54\textwidth]{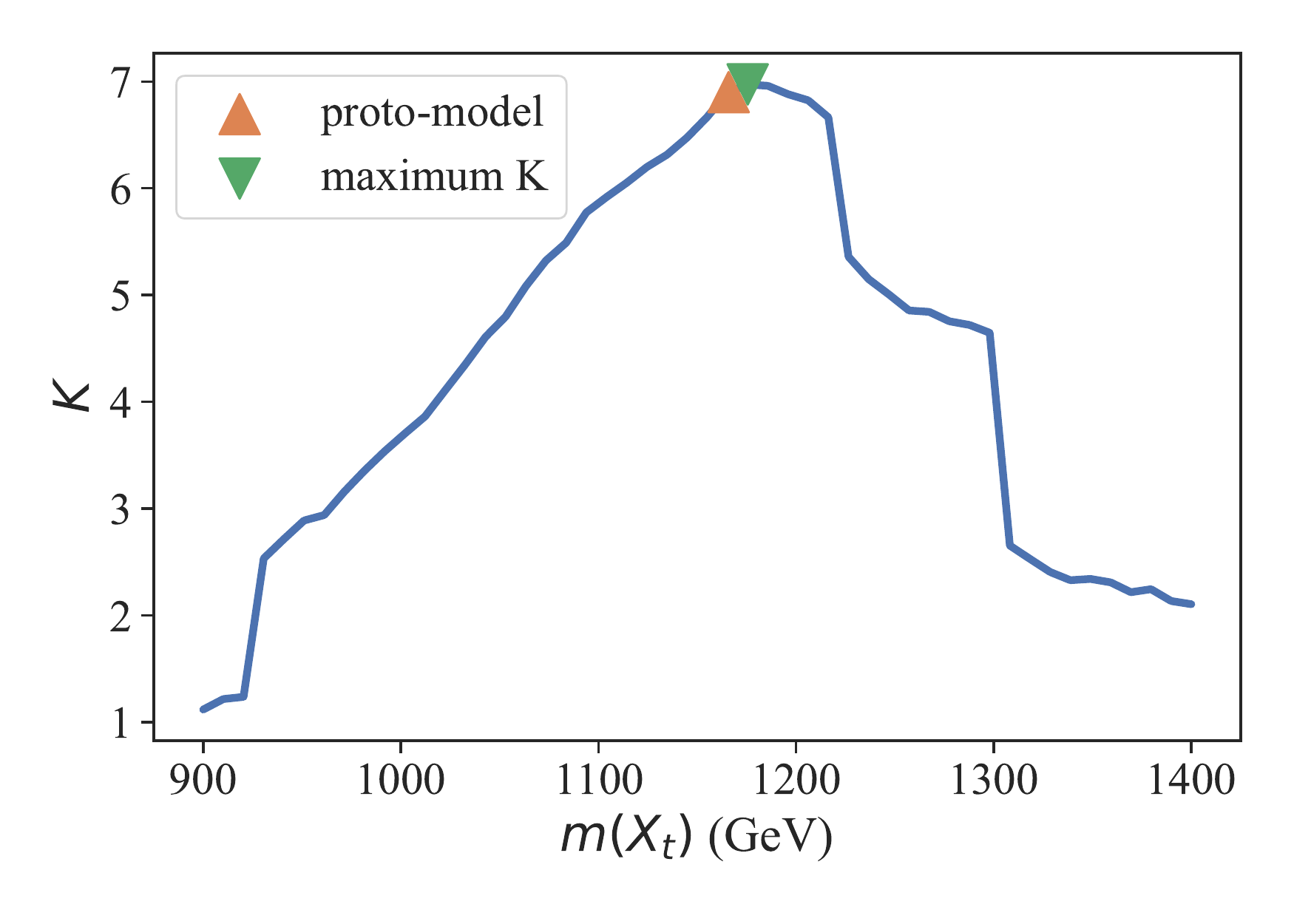}\\
	\includegraphics[width=.54\textwidth]{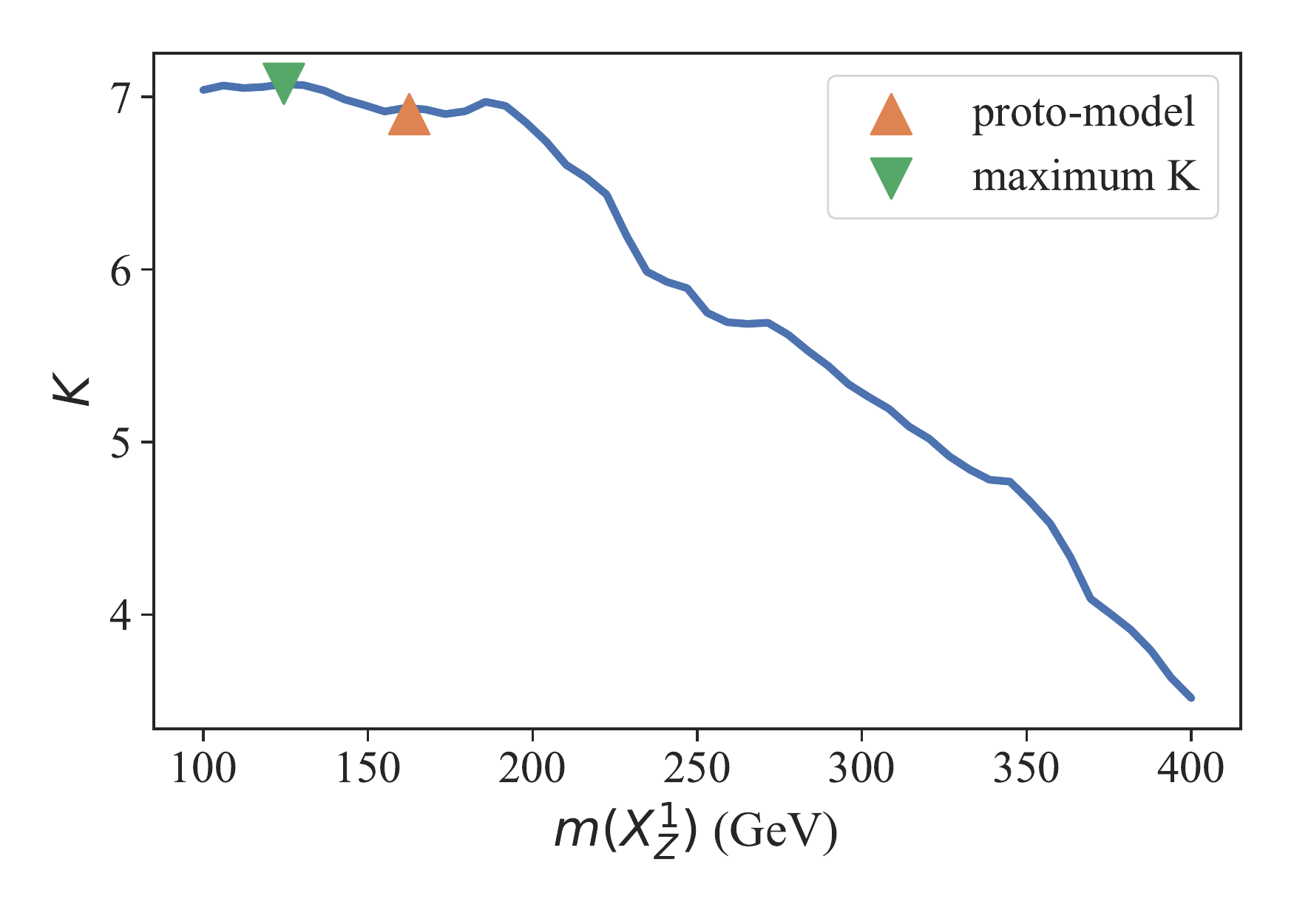}
	\caption{The test statistic \K (blue line) in the vicinity of the highest-score \protomodel, run 9 from Fig.~\ref{fig:winners}, as a function of this \protomodel's masses; on the top left for $X_d$, top right for $X_t$ and on the bottom for $X_Z^1$. The maximum \K in each case is indicated by a green downward triangle, and the \K of the highest-score \protomodel as an orange upward triangle.}
	\label{fig:neighborhood}
\end{center}\end{figure}

\begin{figure}[t!] 
	\hspace*{-6mm}\includegraphics[width=.54\textwidth]{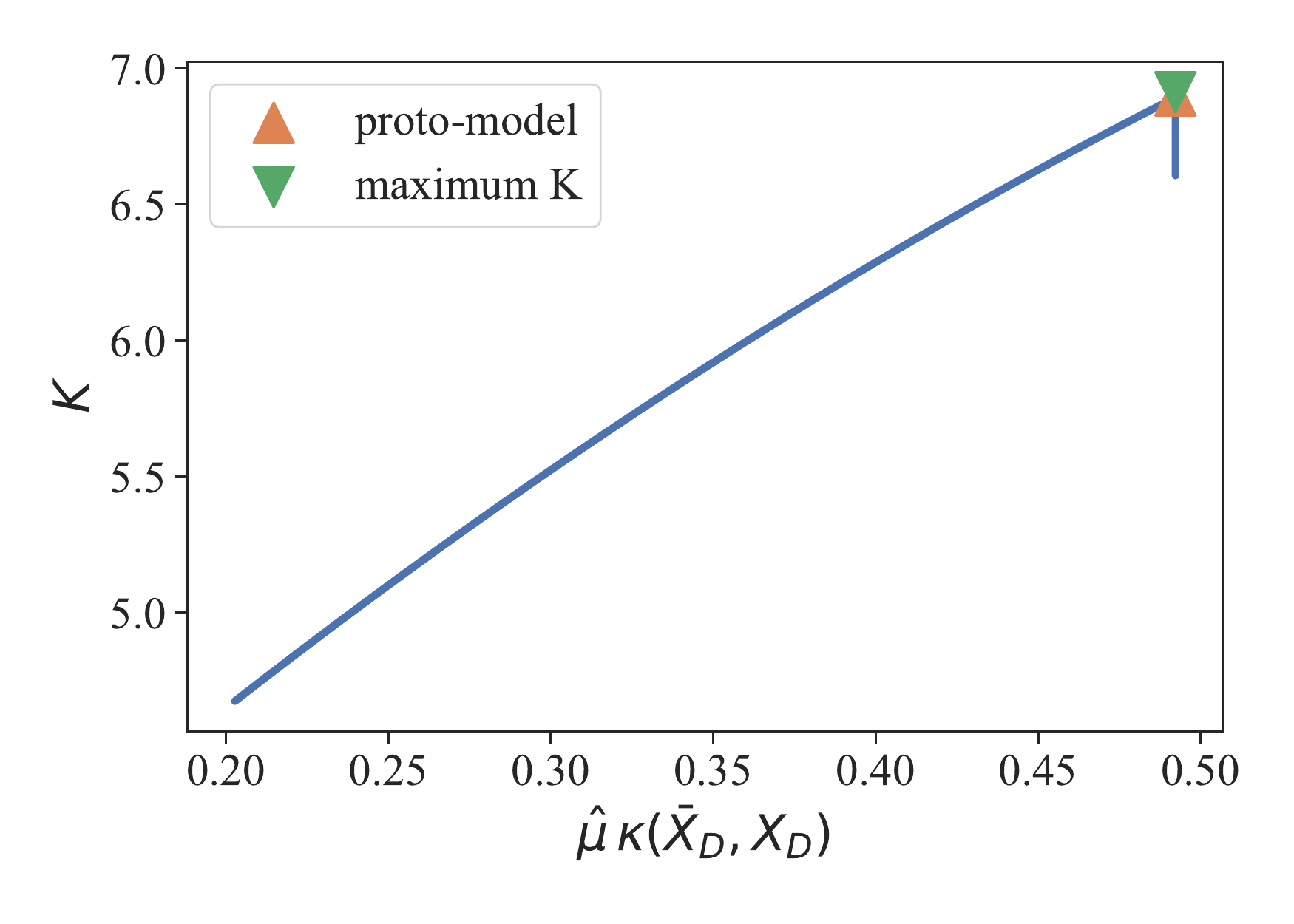}%
	\hspace*{-2mm}\includegraphics[width=.54\textwidth]{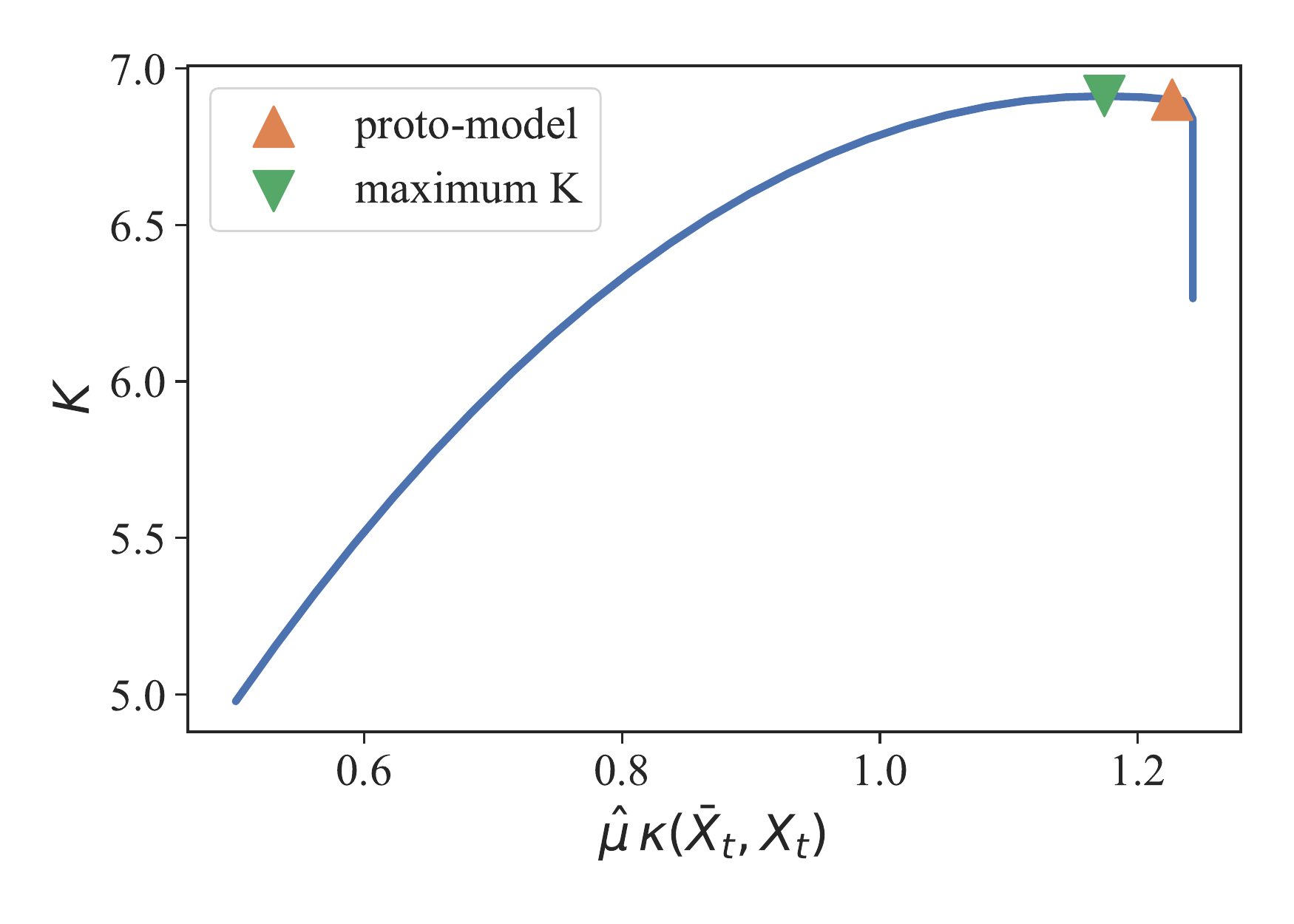}
	\vspace*{-8mm} 
	\caption{As Fig.~\ref{fig:neighborhood} but as a function of the \protomodel's signal strength multipliers. The sharp cut at large values is due to the constraint $\mu < \mu_{max}$ obtained from the critic.} 
	\label{fig:neighborhood_ssm}
\end{figure}

In order to illustrate the convergence of the walker algorithm, we show in Fig.~\ref{fig:neighborhood} 
the behavior of the test statistic close to the highest-score  \protomodel (cf.\ run~9 in Fig.~\ref{fig:winners}) as a function of the masses of the BSM particles ($X_d$, $X_t$ and $X_Z^1$). As we can see, the \protomodel generated by the walker is very close to the maximum of the 1D distributions, showing that the best score obtained during the walk has indeed converged to the (local) maximum.
The same behavior is seen in Fig.~\ref{fig:neighborhood_ssm}, where we display the \K distribution as a function of the signal strength multipliers. All the curves in Fig.~\ref{fig:neighborhood_ssm} display a sharp cut on $\hat{\mu} \kappa$ at large $\kappa$ values. This behavior is due to the limit imposed on the signal strength by the critic. Hence, as $\kappa$ increases, $\hat{\mu}$ must decrease in order to satisfy the constraints on the total signal cross-section, which is proportional to  $\hat{\mu}\kappa$. This explicitly shows the tension between the critic, which limits  $\mu$, and the \protomodel builder, which would tend to increase $\kappa$ in order to maximize \K.

\subsubsection*{Posteriors of Particle Masses}

\begin{figure}[h!]
	\hspace{-2mm}\includegraphics[width=.54\textwidth]{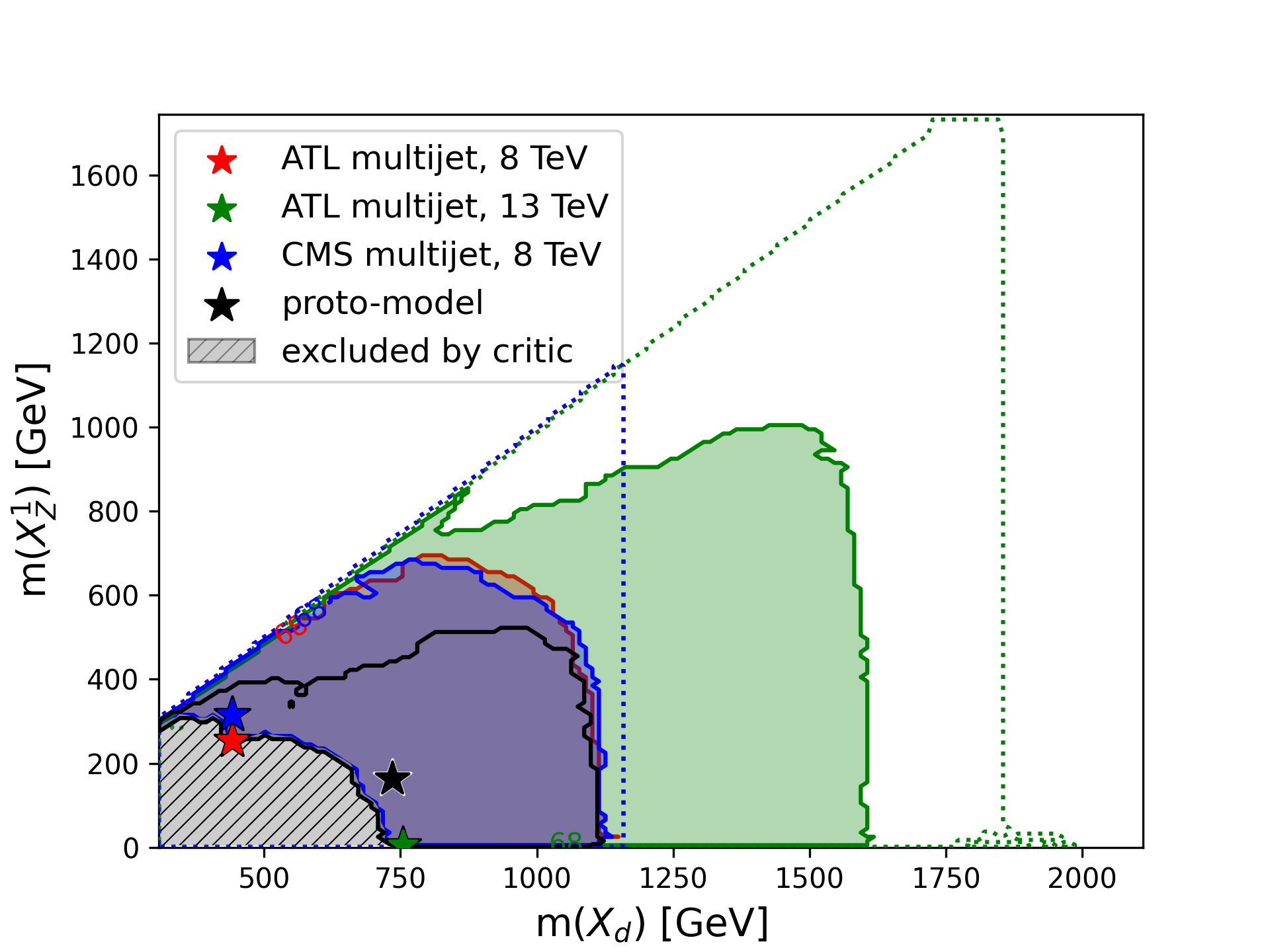}\hspace{-4mm}%
	\includegraphics[width=.54\textwidth]{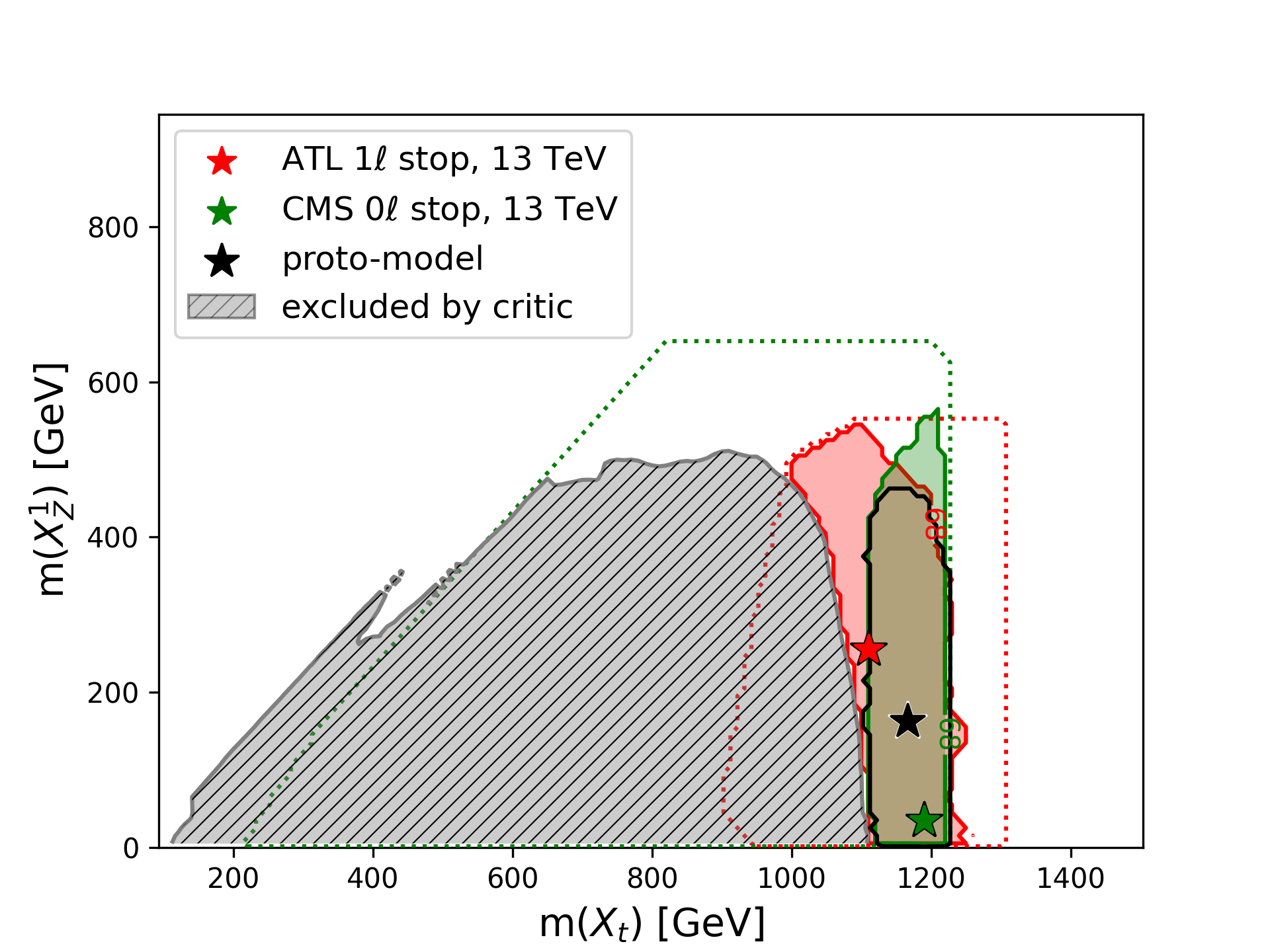} 
	\caption{The 68\% Bayesian credibility regions (coloured areas, solid contours) in the mass vs.\ mass planes of the corresponding simplified model: on the left for  
	($X_d, X_Z^1$), on the right for ($X_t^1, X_Z^1$). 
    The signal strength multipliers are fixed at the values of the highest-score \protomodel with $\K=6.9$. The coloured stars show the preferred points for each individual analysis, while the black stars display the \protomodel parameters. The hatched areas are excluded by the critic. 
    The dashed lines indicate the size of the UL and EM maps provided by the experiments, thus defining the 100\% credibility regions.}
	\label{fig:compatibiity}
\end{figure}

Another interesting question is that of the mutual compatibility of the excesses.
To address this question, we show in Fig.~\ref{fig:compatibiity} the 68\% Bayesian credibility regions~(BCRs) in the relevant mass vs.\ mass planes  for each set of analyses. The BCRs for the jets$+E_T^{\rm miss}$ searches are shown in the top panel by the colored regions. The combined BCR is shown by the solid black curve. As we can see, the dependence on the masses is rather flat making all the BCRs compatible with each other. We also see that the preferred points are at the edge of the region allowed by the critic, which once more displays the tension between the critic and the excesses.
The bottom panel shows the corresponding regions for the searches for stops. As in the previous case, all the regions are compatible.
Note, moreover, the sharp vertical cut-offs at mother masses of 1--2~TeV.
These are directly inherited from the size of the UL and EM maps provided by the experiments (indicated as dotted lines in Fig.~\ref{fig:compatibiity}). For example, the UL map for the stop simplified model in the CMS $0l$ stop search at 13 TeV extends only up to $\approx 1225$ GeV. We will try and remedy such artefacts by producing  larger maps ourselves, similar to what has been done for gluino-squark production in \cite{Khosa:2020zar}. 

\subsection{Overview of Results in the \smo v1.2.4 Database}
\label{sec:anatables}

Tables~\ref{tab:anatab_atlas13}--\ref{tab:anatab_cms8} list the ATLAS and CMS analyses 
for which simplified-model results are included the \smo v1.2.4 database. 
Also listed are the available types of results: observed and/or expected upper limits (UL$_{\rm obs}$, UL$_{\rm exp}$), efficiency maps (EM), and, in the column denoted ``comb.'', what information is available for the combination of signal regions in EM results, if any. No entry means no such information is available.

In the ``short description'', the following abbreviations are used when needed. $\ell$: leptons, SS: same-sign, OS: opposite-sign, SF: same-flavor, EWino: electroweakino, H(bb): Higgs decaying into $b\bar b$. 
Note that all analyses have $E_T^{\rm miss}$ in the final state, which is omitted in the short description for the sake of brevity. The integrated luminosity is in units of fb$^{-1}$.  

We remind the reader that, when only observed ULs are available, no proper likelihood can be constructed; instead only a step function at the 95\% CL limit can be used.

\begin{table}[h] \small 
\begin{tabular}{|l|l|c|c|c|c|c|c|c|}
\hline
{\bf ID} & {\bf Short Description} & {\bf $\mathcal{L}$ [fb$^{-1}$]\!} & {\bf UL$_\mathrm{obs}$} & {\bf UL$_\mathrm{exp}$} & {\bf EM}& {\bf comb.}\\
\hline
\href{https://atlas.web.cern.ch/Atlas/GROUPS/PHYSICS/PAPERS/SUSY-2015-01/}{ATLAS-SUSY-2015-01}~\cite{Aaboud:2016nwl} & 2 b-jets & 3.2 & \checkmark &   &  &   \\
\href{https://atlas.web.cern.ch/Atlas/GROUPS/PHYSICS/PAPERS/SUSY-2015-02/}{ATLAS-SUSY-2015-02}~\cite{Aaboud:2016lwz} & 1$\ell$ stop &3.2 & \checkmark &   & \checkmark&   \\
\href{http://atlas.web.cern.ch/Atlas/GROUPS/PHYSICS/PAPERS/SUSY-2015-06/}{ATLAS-SUSY-2015-06}~\cite{Aaboud:2016zdn} & 0$\ell$ + 2--6 jets  & 3.2 &   &   & \checkmark&   \\
\href{https://atlas.web.cern.ch/Atlas/GROUPS/PHYSICS/PAPERS/SUSY-2015-09/}{ATLAS-SUSY-2015-09}~\cite{Aad:2016tuk} & jets + 2 SS or $\ge 3\,\ell$ &3.2 & \checkmark &   &  &   \\
\href{https://atlas.web.cern.ch/Atlas/GROUPS/PHYSICS/PAPERS/SUSY-2016-07/}{ATLAS-SUSY-2016-07}~\cite{Aaboud:2017vwy} & 0$\ell$ + jets  &36.1 & \checkmark &   & \checkmark&   \\
\href{http://atlas.web.cern.ch/Atlas/GROUPS/PHYSICS/PAPERS/SUSY-2016-14/}{ATLAS-SUSY-2016-14}~\cite{Aaboud:2017dmy} & jets + 2 SS or $\ge 3\,\ell$ & 36.1 & \checkmark &   &  &   \\
\href{https://atlas.web.cern.ch/Atlas/GROUPS/PHYSICS/PAPERS/SUSY-2016-15/}{ATLAS-SUSY-2016-15}~\cite{Aaboud:2017ayj} & 0$\ell$ stop &36.1 & \checkmark &   &  &   \\
\href{https://atlas.web.cern.ch/Atlas/GROUPS/PHYSICS/PAPERS/SUSY-2016-16/}{ATLAS-SUSY-2016-16}~\cite{Aaboud:2017aeu} & 1$\ell$ stop &36.1 & \checkmark &   & \checkmark&   \\
\href{http://atlas.web.cern.ch/Atlas/GROUPS/PHYSICS/PAPERS/SUSY-2016-17/}{ATLAS-SUSY-2016-17}~\cite{Aaboud:2017nfd} & 2 OS leptons  &36.1 & \checkmark &   &  &   \\
\href{https://atlas.web.cern.ch/Atlas/GROUPS/PHYSICS/PAPERS/SUSY-2016-19/}{ATLAS-SUSY-2016-19}~\cite{Aaboud:2018kya} & 2 b-jets + $\tau$'s &36.1 & \checkmark &   &  &   \\
\href{https://atlas.web.cern.ch/Atlas/GROUPS/PHYSICS/PAPERS/SUSY-2016-24/}{ATLAS-SUSY-2016-24}~\cite{Aaboud:2018jiw} & 2--3 $\ell$, EWino &36.1 & \checkmark &   & \checkmark&   \\
\href{https://atlas.web.cern.ch/Atlas/GROUPS/PHYSICS/PAPERS/SUSY-2016-26/}{ATLAS-SUSY-2016-26}~\cite{Aaboud:2018zjf} & $\ge 2$ c-jets  &36.1 & \checkmark &   &  &   \\
\href{https://atlas.web.cern.ch/Atlas/GROUPS/PHYSICS/PAPERS/SUSY-2016-27/}{ATLAS-SUSY-2016-27}~\cite{Aaboud:2018doq} & jets + $\gamma$  &36.1 & \checkmark &   & \checkmark&   \\
\href{https://atlas.web.cern.ch/Atlas/GROUPS/PHYSICS/PAPERS/SUSY-2016-28/}{ATLAS-SUSY-2016-28}~\cite{Aaboud:2017wqg} & 2 b-jets  &36.1 & \checkmark &   &  &   \\
\href{https://atlas.web.cern.ch/Atlas/GROUPS/PHYSICS/PAPERS/SUSY-2016-33/}{ATLAS-SUSY-2016-33}~\cite{Aaboud:2018ujj} & 2 OSSF $\ell$  &36.1 & \checkmark &   &  &   \\
\href{https://atlas.web.cern.ch/Atlas/GROUPS/PHYSICS/PAPERS/SUSY-2017-01/}{ATLAS-SUSY-2017-01}~\cite{Aaboud:2018ngk} & WH(bb), EWino  &36.1 & \checkmark &   &  &   \\
\href{https://atlas.web.cern.ch/Atlas/GROUPS/PHYSICS/PAPERS/SUSY-2017-02/}{ATLAS-SUSY-2017-02}~\cite{Aaboud:2018htj} & 0$\ell$ + jets  &36.1 & \checkmark & \checkmark &  &   \\
\href{https://atlas.web.cern.ch/Atlas/GROUPS/PHYSICS/PAPERS/SUSY-2017-03/}{ATLAS-SUSY-2017-03}~\cite{Aaboud:2018sua} & 2--3 leptons, EWino &36.1 & \checkmark &   &  &   \\
\href{https://atlas.web.cern.ch/Atlas/GROUPS/PHYSICS/PAPERS/SUSY-2018-04/}{ATLAS-SUSY-2018-04}~\cite{Aad:2019byo} & 2 hadronic taus  &139.0 & \checkmark &   & \checkmark& JSON \\
\href{https://atlas.web.cern.ch/Atlas/GROUPS/PHYSICS/PAPERS/SUSY-2018-06/}{ATLAS-SUSY-2018-06}~\cite{Aad:2019vvi} & 3 leptons, EWino &139.0 & \checkmark & \checkmark &  &   \\
\href{https://atlas.web.cern.ch/Atlas/GROUPS/PHYSICS/PAPERS/SUSY-2018-31/}{ATLAS-SUSY-2018-31}~\cite{Aad:2019pfy} & 2b + 2H(bb)  &139.0 & \checkmark &   & \checkmark& JSON \\
\href{https://atlas.web.cern.ch/Atlas/GROUPS/PHYSICS/PAPERS/SUSY-2018-32/}{ATLAS-SUSY-2018-32}~\cite{Aad:2019vnb} & 2 OS leptons  &139.0 & \checkmark &   &  &   \\
\href{https://atlas.web.cern.ch/Atlas/GROUPS/PHYSICS/PAPERS/SUSY-2019-08/}{ATLAS-SUSY-2019-08}~\cite{Aad:2019vvf} & 1$\ell$ + higgs   &139.0 & \checkmark &   & \checkmark& JSON \\
\hline
\end{tabular}

\caption{List of ATLAS Run~2 analyses and their types of results in the \smo v1.2.4 database.}
\label{tab:anatab_atlas13}
\end{table}

\begin{table}[h] \small \hspace*{-2mm}%
\begin{tabular}{|l|l|c|c|c|c|c|c|c|}
\hline
{\bf ID} & {\bf Short Description} & {\bf $\mathcal{L}$ [fb$^{-1}$]\!} & {\bf UL$_\mathrm{obs}$} & {\bf UL$_\mathrm{exp}$} & {\bf EM} & {\bf comb.}\\
\hline
\href{http://cms-results.web.cern.ch/cms-results/public-results/preliminary-results/EXO-16-036/index.html}{CMS-PAS-EXO-16-036}~\cite{CMS-PAS-EXO-16-036} & HSCP &12.9 & \checkmark &   & \checkmark &   \\
\href{http://cms-results.web.cern.ch/cms-results/public-results/preliminary-results/SUS-16-052/index.html}{CMS-PAS-SUS-16-052}~\cite{CMS-PAS-SUS-16-052} & ISR jet + soft $\ell$ & 35.9 & \checkmark &   & \checkmark & Cov.  \\
\href{https://cms-results.web.cern.ch/cms-results/public-results/publications/SUS-16-009/}{CMS-SUS-16-009}~\cite{Khachatryan:2017rhw} & 0$\ell$ + jets, top tagging & 2.3 & \checkmark & \checkmark &   &   \\
\href{http://cms-results.web.cern.ch/cms-results/public-results/publications/SUS-16-032/index.html}{CMS-SUS-16-032}~\cite{Sirunyan:2017kiw} & 2 b- or 2 c-jets  & 35.9 & \checkmark &   &   &   \\
\href{http://cms-results.web.cern.ch/cms-results/public-results/publications/SUS-16-033/index.html}{CMS-SUS-16-033}~\cite{Sirunyan:2017cwe} & 0$\ell$ + jets   & 35.9 & \checkmark & \checkmark & \checkmark &   \\
\href{http://cms-results.web.cern.ch/cms-results/public-results/publications/SUS-16-034/index.html}{CMS-SUS-16-034}~\cite{Sirunyan:2017qaj} & 2 OSSF leptons & 35.9 & \checkmark &   &   &   \\
\href{http://cms-results.web.cern.ch/cms-results/public-results/publications/SUS-16-035/index.html}{CMS-SUS-16-035}~\cite{Sirunyan:2017uyt} & 2 SS leptons & 35.9 & \checkmark &   &   &   \\
\href{http://cms-results.web.cern.ch/cms-results/public-results/publications/SUS-16-036/index.html}{CMS-SUS-16-036}~\cite{Sirunyan:2017kqq} & 0$\ell$ + jets   & 35.9 & \checkmark & \checkmark &   &   \\
\href{http://cms-results.web.cern.ch/cms-results/public-results/publications/SUS-16-037/index.html}{CMS-SUS-16-037}~\cite{Sirunyan:2017fsj} & 1$\ell$ + jets with MJ  & 35.9 & \checkmark &   &   &   \\
\href{http://cms-results.web.cern.ch/cms-results/public-results/publications/SUS-16-039/index.html}{CMS-SUS-16-039}~\cite{Sirunyan:2017lae} & 2--3$\ell$, EWino & 35.9 & \checkmark &   &   &   \\
\href{http://cms-results.web.cern.ch/cms-results/public-results/publications/SUS-16-041/index.html}{CMS-SUS-16-041}~\cite{Sirunyan:2017hvp} & jets + $\geq 3\ell$ & 35.9 & \checkmark &   &   &   \\
\href{http://cms-results.web.cern.ch/cms-results/public-results/publications/SUS-16-042/index.html}{CMS-SUS-16-042}~\cite{Sirunyan:2017mrs} & 1$\ell$ + jets   & 35.9 & \checkmark &   &   &   \\
\href{http://cms-results.web.cern.ch/cms-results/public-results/publications/SUS-16-043/index.html}{CMS-SUS-16-043}~\cite{Sirunyan:2017zss} & WH(bb), EWino & 35.9 & \checkmark &   &   &   \\
\href{http://cms-results.web.cern.ch/cms-results/public-results/publications/SUS-16-045/index.html}{CMS-SUS-16-045}~\cite{Sirunyan:2017eie} & jets + H $\rightarrow$ $\gamma\gamma$  & 35.9 & \checkmark &   &   &   \\
\href{http://cms-results.web.cern.ch/cms-results/public-results/publications/SUS-16-046/index.html}{CMS-SUS-16-046}~\cite{Sirunyan:2017nyt} & high-$p_T$ $\gamma$  & 35.9 & \checkmark &   &   &   \\
\href{http://cms-results.web.cern.ch/cms-results/public-results/publications/SUS-16-047/index.html}{CMS-SUS-16-047}~\cite{Sirunyan:2017yse} & $\gamma$ + jets, high $H_T$ & 35.9 & \checkmark &   &   &   \\
\href{http://cms-results.web.cern.ch/cms-results/public-results/publications/SUS-16-049/index.html}{CMS-SUS-16-049}~\cite{Sirunyan:2017wif} & $0\ell$ stop & 35.9 & \checkmark & \checkmark &   &   \\
\href{http://cms-results.web.cern.ch/cms-results/public-results/publications/SUS-16-050/index.html}{CMS-SUS-16-050}~\cite{Sirunyan:2017pjw} & $0\ell$ stop, $m_{T2}$ & 35.9 & \checkmark & \checkmark &   &   \\
\href{http://cms-results.web.cern.ch/cms-results/public-results/publications/SUS-16-051/index.html}{CMS-SUS-16-051}~\cite{Sirunyan:2017xse} & $1\ell$ stop & 35.9 & \checkmark & \checkmark &   &   \\
\href{http://cms-results.web.cern.ch/cms-results/public-results/publications/SUS-17-001/index.html}{CMS-SUS-17-001}~\cite{Sirunyan:2017leh} & $2\ell$ stop  & 35.9 & \checkmark &   &   &   \\
\href{https://cms-results.web.cern.ch/cms-results/public-results/publications/SUS-17-003/}{CMS-SUS-17-003}~\cite{Sirunyan:2018vig} & 2 taus  & 35.9 & \checkmark &   &   &   \\
\href{http://cms-results.web.cern.ch/cms-results/public-results/publications/SUS-17-004/index.html}{CMS-SUS-17-004}~\cite{Sirunyan:2018ubx} & EWino combination & 35.9 & \checkmark &   &   &   \\
\href{https://cms-results.web.cern.ch/cms-results/public-results/publications/SUS-17-005/}{CMS-SUS-17-005}~\cite{Sirunyan:2018omt} & $1\ell$ stop, soft & 35.9 & \checkmark & \checkmark &   &   \\
\href{https://cms-results.web.cern.ch/cms-results/public-results/publications/SUS-17-006/}{CMS-SUS-17-006}~\cite{Sirunyan:2017bsh} & jets + boosted H(bb)  & 35.9 & \checkmark & \checkmark &   &   \\
\href{https://cms-results.web.cern.ch/cms-results/public-results/publications/SUS-17-009/}{CMS-SUS-17-009}~\cite{Sirunyan:2018nwe} & 2 OSSF leptons  & 35.9 & \checkmark & \checkmark &   &   \\
\href{http://cms-results.web.cern.ch/cms-results/public-results/publications/SUS-17-010}{CMS-SUS-17-010}~\cite{Sirunyan:2018lul} & 2$\ell$ EWino, stop & 35.9 & \checkmark & \checkmark &   &   \\
\href{https://cms-results.web.cern.ch/cms-results/public-results/publications/SUS-18-002/}{CMS-SUS-18-002}~\cite{Sirunyan:2019hzr} & $\gamma$ + (b-)jets & 35.9 & \checkmark & \checkmark &   &   \\
\href{http://cms-results.web.cern.ch/cms-results/public-results/publications/SUS-19-006/index.html}{CMS-SUS-19-006}~\cite{Sirunyan:2019ctn} & $0\ell$ + jets, MHT &137.0 & \checkmark & \checkmark &   &   \\
\hline
\end{tabular}

\caption{List of CMS Run~2 analyses and their types of results in the \smo v1.2.4 database. In the last column, ``Cov.'' stands for covariance matrix.}
\label{tab:anatab_cms13}
\end{table}

\begin{table}[h] \small 
\begin{tabular}{|l|l|c|c|c|c|c|c|c|}
\hline
{\bf ID} & {\bf Short Description} & {\bf $\mathcal{L}$ [fb$^{-1}$]\!} & {\bf UL$_\mathrm{obs}$} & {\bf UL$_\mathrm{exp}$} & {\bf EM}\\
\hline
\href{https://atlas.web.cern.ch/Atlas/GROUPS/PHYSICS/CONFNOTES/ATLAS-CONF-2013-007/}{ATLAS-CONF-2013-007}~\cite{ATLAS-CONF-2013-007} & 2 SS $\ell$ + (b-)jets  & 20.7 & \checkmark &   &   \\
\href{https://atlas.web.cern.ch/Atlas/GROUPS/PHYSICS/CONFNOTES/ATLAS-CONF-2013-061/}{ATLAS-CONF-2013-061}~\cite{ATLAS-CONF-2013-061} &  0--1$\ell$ + 4--7 ($\geq3$ b-)jets  & 20.1 & \checkmark &   &   \\
\href{https://atlas.web.cern.ch/Atlas/GROUPS/PHYSICS/CONFNOTES/ATLAS-CONF-2013-089/}{ATLAS-CONF-2013-089}~\cite{ATLAS-CONF-2013-089} & 2$\ell$ + jets  & 20.3 & \checkmark &   &   \\ 
\href{https://atlas.web.cern.ch/Atlas/GROUPS/PHYSICS/PAPERS/SUSY-2013-02/}{ATLAS-SUSY-2013-02}~\cite{Aad:2014wea} & 0$\ell$ + 2--6 jets  & 20.3 & \checkmark &   & \checkmark \\
\href{https://atlas.web.cern.ch/Atlas/GROUPS/PHYSICS/PAPERS/SUSY-2013-04/}{ATLAS-SUSY-2013-04}~\cite{Aad:2013wta} & 0$\ell$ + 7--10 jets  & 20.3 & \checkmark &   & \checkmark \\
\href{https://atlas.web.cern.ch/Atlas/GROUPS/PHYSICS/PAPERS/SUSY-2013-05/}{ATLAS-SUSY-2013-05}~\cite{Aad:2013ija} & 0$\ell$ + 2 b-jets  &20.1 & \checkmark &   & \checkmark \\
\href{https://atlas.web.cern.ch/Atlas/GROUPS/PHYSICS/PAPERS/SUSY-2013-08/}{ATLAS-SUSY-2013-08}~\cite{Aad:2014mha} & Z + b-jets  & 20.3 & \checkmark &   &   \\
\href{https://atlas.web.cern.ch/Atlas/GROUPS/PHYSICS/PAPERS/SUSY-2013-09/}{ATLAS-SUSY-2013-09}~\cite{Aad:2014pda} & jets + 2 SS or $\geq3\ell$  & 20.3 & \checkmark &   &   \\
\href{https://atlas.web.cern.ch/Atlas/GROUPS/PHYSICS/PAPERS/SUSY-2013-11/}{ATLAS-SUSY-2013-11}~\cite{Aad:2014vma} & 2$\ell$ ($e,\mu$)  & 20.3 & \checkmark &   & \checkmark \\
\href{https://atlas.web.cern.ch/Atlas/GROUPS/PHYSICS/PAPERS/SUSY-2013-12/}{ATLAS-SUSY-2013-12}~\cite{Aad:2014nua} & 3$\ell$ ($e,\mu,\tau$)  & 20.3 & \checkmark &   &   \\
\href{https://atlas.web.cern.ch/Atlas/GROUPS/PHYSICS/PAPERS/SUSY-2013-15/}{ATLAS-SUSY-2013-15}~\cite{Aad:2014kra} & 1$\ell$ + 4 (1 b-)jets  & 20.3 & \checkmark &   & \checkmark \\
\href{https://atlas.web.cern.ch/Atlas/GROUPS/PHYSICS/PAPERS/SUSY-2013-16/}{ATLAS-SUSY-2013-16}~\cite{Aad:2014bva} & 0$\ell$ + 6 (2 b-)jets  &20.1 & \checkmark &   & \checkmark \\
\href{https://atlas.web.cern.ch/Atlas/GROUPS/PHYSICS/PAPERS/SUSY-2013-18/}{ATLAS-SUSY-2013-18}~\cite{Aad:2014lra} & 0-1$\ell$ + $\geq 3$ b-jets  &20.1 & \checkmark &   & \checkmark \\
\href{https://atlas.web.cern.ch/Atlas/GROUPS/PHYSICS/PAPERS/SUSY-2013-19/}{ATLAS-SUSY-2013-19}~\cite{Aad:2014qaa} & 2 OS $\ell$ + (b-)jets   & 20.3 & \checkmark &   &   \\
\href{https://atlas.web.cern.ch/Atlas/GROUPS/PHYSICS/PAPERS/SUSY-2013-21/}{ATLAS-SUSY-2013-21}~\cite{Aad:2014nra} & monojet or c-jet  & 20.3 &   &   & \checkmark \\
\href{https://atlas.web.cern.ch/Atlas/GROUPS/PHYSICS/PAPERS/SUSY-2013-23/}{ATLAS-SUSY-2013-23}~\cite{Aad:2015jqa} & WH ($1\ell+ bb$ or $\gamma\gamma$)   & 20.3 & \checkmark &   &   \\
\href{https://atlas.web.cern.ch/Atlas/GROUPS/PHYSICS/PAPERS/SUSY-2014-03/}{ATLAS-SUSY-2014-03}~\cite{Aad:2015gna} & $\geq 2$(c-)jets  & 20.3 &   &   & \checkmark \\
\hline
\end{tabular}

\caption{List of ATLAS Run~1 analyses and their types of results in the \smo v1.2.4 database. The column ``comb.'' is omitted as no information on signal region correlations is available for any of the analyses.}
\label{tab:anatab_atlas8}
\end{table}

\begin{table}[h] \small 
\begin{tabular}{|l|l|c|c|c|c|c|c|c|}
\hline
{\bf ID} & {\bf Short Description} & {\bf $\mathcal{L}$ [fb$^{-1}$] } & {\bf UL$_\mathrm{obs}$} & {\bf UL$_\mathrm{exp}$} & {\bf EM}\\
\hline
\href{http://cms-results.web.cern.ch/cms-results/public-results/publications/EXO-12-026/index.html}{CMS-EXO-12-026}~\cite{Chatrchyan:2013oca} & HSCP &18.8 & \checkmark &   &   \\
\href{http://cms-results.web.cern.ch/cms-results/public-results/publications/EXO-13-006/index.html}{CMS-EXO-13-006}~\cite{Khachatryan:2015lla} & HSCP &18.8 &   &   & \checkmark \\
\href{https://twiki.cern.ch/twiki/bin/view/CMSPublic/PhysicsResultsSUS13015}{CMS-PAS-SUS-13-015}~\cite{CMS-PAS-SUS-13-015} & $\ge 5$(b-)jets, top tag & 19.4 &   &   & \checkmark \\
\href{https://twiki.cern.ch/twiki/bin/view/CMSPublic/PhysicsResultsSUS13016}{CMS-PAS-SUS-13-016}~\cite{CMS-PAS-SUS-13-016} & 2 OS $\ell$ + (b-)jets &19.7 & \checkmark &   & \checkmark \\
\href{https://twiki.cern.ch/twiki/bin/view/CMSPublic/PhysicsResultsSUS13018}{CMS-PAS-SUS-13-018}~\cite{CMS-PAS-SUS-13-018} & 1--2 b-jets, $M_{CT}$ &19.4 & \checkmark &   &   \\
\href{https://twiki.cern.ch/twiki/bin/view/CMSPublic/PhysicsResultsSUS13023}{CMS-PAS-SUS-13-023}~\cite{CMS-PAS-SUS-13-023} & $0\ell$ stop &18.9 & \checkmark &   &   \\
\href{https://twiki.cern.ch/twiki/bin/view/CMSPublic/PhysicsResultsSUS12024}{CMS-SUS-12-024}~\cite{Chatrchyan:2013wxa} & $0\ell$ + $\ge 3$\,(b-)jets &19.4 & \checkmark &   & \checkmark \\
\href{https://twiki.cern.ch/twiki/bin/view/CMSPublic/PhysicsResultsSUS12028}{CMS-SUS-12-028}~\cite{Chatrchyan:2013mys} & $0\ell$ + (b-)jets, $\alpha_T$ &11.7 & \checkmark & \checkmark &   \\
\href{https://twiki.cern.ch/twiki/bin/view/CMSPublic/PhysicsResultsSUS13002}{CMS-SUS-13-002}~\cite{Chatrchyan:2014aea} & $\ge 3\ell$ (+jets)  &19.5 & \checkmark & \checkmark &   \\
\href{https://twiki.cern.ch/twiki/bin/view/CMSPublic/PhysicsResultsSUS13004}{CMS-SUS-13-004}~\cite{Khachatryan:2015pwa} & $\ge 1$ b-jet, Razor &19.3 & \checkmark &   &   \\
\href{https://twiki.cern.ch/twiki/bin/view/CMSPublic/PhysicsResultsSUS13006}{CMS-SUS-13-006}~\cite{Khachatryan:2014qwa} & multi-$\ell$ EWino &19.5 & \checkmark &   &   \\
\href{https://twiki.cern.ch/twiki/bin/view/CMSPublic/PhysicsResultsSUS13007}{CMS-SUS-13-007}~\cite{Chatrchyan:2013iqa} & $1\ell$ + $\ge 2$ b-jets &19.3 & \checkmark &   & \checkmark \\
\href{https://twiki.cern.ch/twiki/bin/view/CMSPublic/PhysicsResultsSUS13011}{CMS-SUS-13-011}~\cite{Chatrchyan:2013xna} & $1\ell$ + $\ge 4$ (1b-)jets + $\not{\!\!E}_T$ &19.5 & \checkmark &   & \checkmark \\
\href{https://twiki.cern.ch/twiki/bin/view/CMSPublic/PhysicsResultsSUS13012}{CMS-SUS-13-012}~\cite{Chatrchyan:2014lfa} & $0\ell$ + 3--8 jets, MHT &19.5 & \checkmark & \checkmark & \checkmark \\
\href{https://twiki.cern.ch/twiki/bin/view/CMSPublic/PhysicsResultsSUS13013}{CMS-SUS-13-013}~\cite{Chatrchyan:2013fea} & 2 SS $\ell$ + (b-)jets  &19.5 & \checkmark & \checkmark & \checkmark \\
\href{https://twiki.cern.ch/twiki/bin/view/CMSPublic/PhysicsResultsSUS13019}{CMS-SUS-13-019}~\cite{Khachatryan:2015vra} & $\ge 2$ jets, $M_{\rm T2}$ &19.5 & \checkmark &   &   \\
\href{https://twiki.cern.ch/twiki/bin/view/CMSPublic/PhysicsResultsSUS14010}{CMS-SUS-14-010}~\cite{CMS:2014dpa} & b-jets + 4 $W$s (0--4$\ell$) &19.5 & \checkmark & \checkmark &   \\
\href{https://twiki.cern.ch/twiki/bin/view/CMSPublic/PhysicsResultsSUS14021}{CMS-SUS-14-021}~\cite{Khachatryan:2015pot} & ISR jet + 1--2 soft $\ell$ & 19.7 & \checkmark & \checkmark &   \\
\hline
\end{tabular}

\caption{List of CMS Run~1 analyses and their types of results in the \smo v1.2.4 database. The column ``comb.'' is omitted as no information on signal region correlations is available for any of the analyses.}
\label{tab:anatab_cms8}
\end{table}

\clearpage

\end{document}